\documentclass[usegraphicx,usenatbib]{mn2e}

\title[Leo~II]{Leo~II Group: decoupled cores of NGC~3607 and NGC~3608
\thanks{Based on observations collected with the 6m
telescope of the Special Astrophysical Observatory (SAO) of the
Russian Academy of Sciences (RAS) which is operated under the
financial support of Science Department of Russia (registration number
01-43) and on data from the Isaac Newton Group (ING) and Hubble Space
Telescope (HST) Archives.}
}
\author[Afanasiev \& Sil'chenko] 
{V.~L.~Afanasiev$^1$ and O.~K.~Sil'chenko$^{2,3}$ 
\thanks{For offprints address to O. K. Sil'chenko, E-mail: olga@sai.msu.su}\\
$^1$Special Astrophysical Observatory, Nizhnij Arkhyz, 369167 Russia\\
$^2$Sternberg Astronomical Institute, University av. 13,
Moscow 119992, Russia\\
$^3$UK Astronomy Data Centre, Guest Investigator
}

\begin{document}
\date{Received / Accepted }
\maketitle

\begin{abstract}
The kinematics, structure, and stellar population properties
in the centers of two brightest
early-type galaxies of the Leo~II group, NGC~3607 and NGC~3608,
are studied by means of integral-field spectroscopy. The
kinematically distinct areas in the centers of these galaxies,
with radii of $6\arcsec$
and $5\arcsec$ respectively, are found also to be chemically distinct.
These stellar structures are characterized by enhanced magnesium-line strength
in the integrated spectra. 
However, we have not found any mean stellar age differences between the decoupled
cores and their outskirts. Analysis of two-dimensional line-of-sight
velocity fields reveals systematic turns of kinematical major axes near the
nuclei of both galaxies; in NGC~3608 the ionized gas rotates in the orthogonal
plane with respect to the stellar component rotation. By taking into
account some morphological features, we conclude that both NGC~3607 and
NGC~3608 have large triaxial stellar spheroids. We argue that
the magnesium-enhanced cores are not circumnuclear disks; 
instead they resemble rather compact triaxial
structures which may be a cause of formation of polar disks around them
-- a gaseous one in NGC~3608 and a stellar-gaseous one in NGC~3607.
In the latter galaxy the star formation is perhaps still proceeding over the polar disk.
\end{abstract}

\section{Introduction}

There exists now a certain contradiction between a homogeneous red-colour
appearance of nearby massive early-type galaxies implying very ancient
epoch of the main star formation and prescriptions of the hierarchical
concept of galaxy formation requiring quite recent merger events. The recent
merging of disk galaxies possessing significant amount of gas must be followed
by some secondary, preferably circumnuclear star formation. Galaxy groups represent
perhaps the best places where consequences of the external-driven evolution 
of early-type galaxies may be found because dense environments and moderate 
(with respect to clusters') velocity dispersions are favourable for mergers
and tidal interactions.

Recently we have considered Leo~I group and its members NGC~3379, NGC~3384, 
and NGC~3368 \citep{we2003}. A combined analysis of the stellar
and gaseous kinematics and of stellar population properties in the centers
of the galaxies has revealed signatures of synchronous secular evolution.
In the centers of the galaxies the inner rotation axes are aligned, despite 
the different orientations of the outer galactic bodies, and the mean stellar 
age estimates evidence for quasi-simultaneous star formation bursts about 
3 Gyr ago.  However, the Leo~I group is somewhat unique because it possesses 
a supergiant intergalactic H~I cloud of $1.7 \cdot 10^9\, M_{\odot}$ 
\citep{leoh1_2,leoh1_3}. Its shape is a clumpy ring with a radius of
$\sim 100$ kpc encircling the galaxy pair NGC~3379/NGC~3384, and 
just the spatial orientation of this ring defines the alignment of the inner rotation
axes of the three galaxies. So it seems probable that the secular evolution
of the Leo~I early-type galaxies is governed by the interaction
of every galaxy with the intergalactic H~I ring and not by tidal
interactions between the galaxies. In the present paper we consider central
early-type galaxies of the Leo~II group lacking significant masses
of neutral hydrogen. It is interesting to look for signatures of the
secular evolution in the circumnuclear parts of these galaxies.
The Leo~II group contains 16 galaxies brighter than $B_T \approx 16$,
according to \citet{nog}. From those, only 5 are of
early type, namely, S0 or ellipticals;  three early-type galaxies,
NGC~3605 (compact dwarf elliptical), NGC~3607, and NGC~3608 are located
in the very center of the group within some 60 kpc area.
The velocity dispersion of the whole group is rather moderate,
417~km/s according to \citet{statgr}.

\begin{table}
\caption[ ] {Global parameters of the galaxies}
\begin{flushleft}
\begin{tabular}{lcc}
\hline\noalign{\smallskip}
NGC & 3607 & 3608  \\
\hline
Type (NED$^1$) & SA(s)0* & E2   \\
$R_{25}$, kpc (LEDA$^2$) & 15.2 & 10.5 \\
$B_T^0$ (RC3$^3$) & 10.79 & 11.69  \\
$M_B$(LEDA)  & --19.84 & --19.46  \\
$(B-V)_T^0$ (RC3) & 0.92 & 0.93  \\
$(U-B)_T^0$ (RC3) & 0.49 & 0.40 \\
$V_r $ (NED), $\mbox{km} \cdot \mbox{s}^{-1}$ &   935 &
      1253   \\
Distance, Mpc  & \multicolumn{2}{c}{$23^4$} \\
Inclination (LEDA) & $34^{\circ}$ & $54^{\circ}$  \\
{\it PA}$_{phot}$ (LEDA) & $120^{\circ}$ & $75^{\circ}$  \\
$\sigma _*$, $\mbox{km} \cdot \mbox{s}^{-1}$(LEDA) & 217 & 204\\
\hline
\multicolumn{3}{l}{$^1$\rule{0pt}{11pt}\footnotesize
NASA/IPAC Extragalactic Database}\\
\multicolumn{3}{l}{$^2$\rule{0pt}{11pt}\footnotesize
Lyon-Meudon Extragalactic Database}\\
\multicolumn{3}{l}{$^3$\rule{0pt}{11pt}\footnotesize
Third Reference Catalogue of Bright Galaxies}\\
\multicolumn{3}{l}{$^4$\rule{0pt}{11pt}\footnotesize
\citet{sbfdist}}
\end{tabular}
\end{flushleft}
\end{table}

\begin{figure*}
\centering
  \includegraphics[width=17cm]{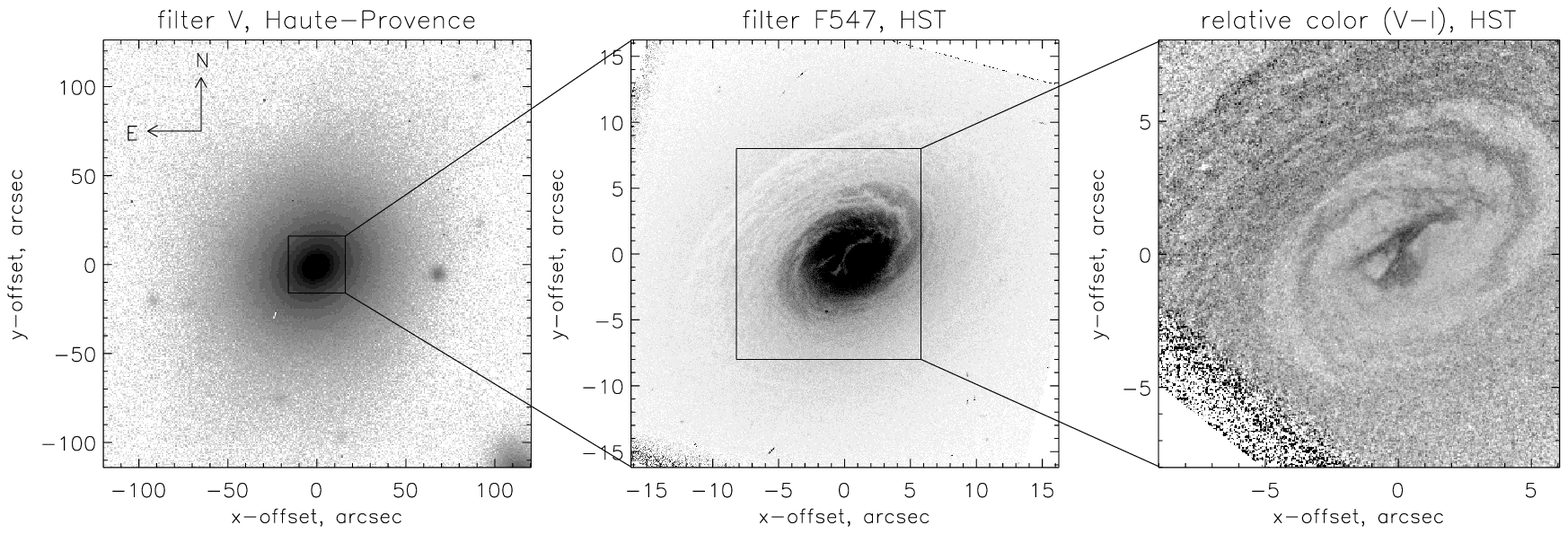}
  \caption{The photometric maps for NGC~3607:
  {\it left --} large-scale image through the filter V (the Haute-Provence
  data from the HYPERLEDA database), intensity is linearly gray-scaled,
  {\it middle --} the HST/PC frame through the filter F547M, the same,
  {\it right --}  the area corresponding to the field of view of the Multi-Pupil Fiber 
  Spectrograph, with the uncalibrated colour distribution obtained by dividing
  the HST/PC frames through the filters F814W and F555W by each other
  and taking 2.5 logarithms, darker means redder.}
  \label{maps3607}
\end{figure*}

\begin{figure*}
\centering
  \includegraphics[width=17cm]{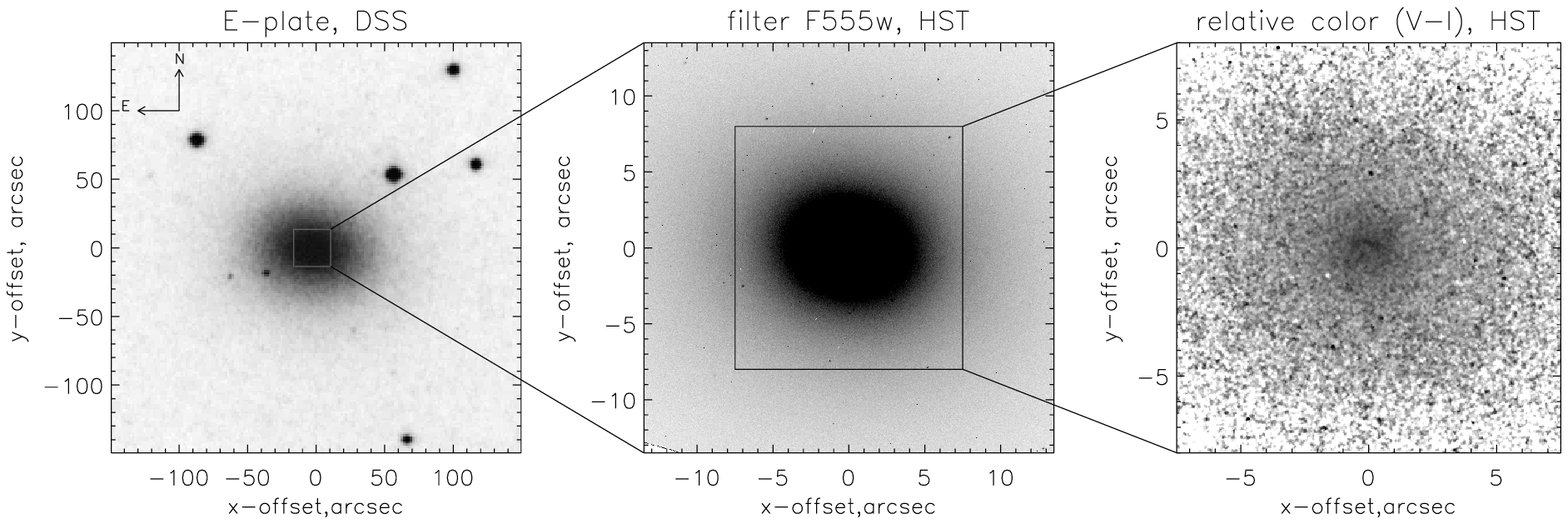}
  \caption{The photometric maps for NGC~3608:
  {\it left --} a large-scale red image from DSS, intensity is linearly
  gray-scaled,
  {\it middle --} the HST/PC frame through the filter F555W, the same
  {\it right --}  the area corresponding to the field of view of  the Multi-Pupil Fiber 
  Spectrograph, with the uncalibrated colour distribution obtained by dividing
  the HST/PC frames through the filters F814W and F555W by each other
  and taking 2.5 logarithms, darker means redder.}
  \label{maps3608}
\end{figure*}

The main characteristics of the galaxies to be considered in the
present paper are given
in Table~1. NGC~3607 is a giant lenticular galaxy located in the
center of the Leo~II group. Hot X-ray gas is detected inside this galaxy
\citep{xrayasca}, as well as around it \citep*{xray82} --
all three galaxies in the center of the group are embedded into
the common envelope of hot gas. An X-ray atlas of \citet{xray02}
demonstrates two maxima of the hot gas concentration -- one at
NGC~3607 and another at NGC~3608 -- which fact forces
authors to suggest a recent assembly of the Leo~II group.
Also NGC~3607, though of early type, is known to have
a rather extended  ionized-gas disk \citep{shields,
macetal} with a radius of $\sim 15\arcsec$ ($\sim 1.5$ kpc) containing
a broad dust ring between $R=8\farcs 4$ and $R=13\farcs 2$ 
\citep{n3607gas}. Central structure of the neighbouring low-luminosity
elliptical galaxy NGC~3608 is also very interesting: it has
a counterrotating core. From their long-slit cross-section taken along
the major axis of NGC~3608 \citet{js88} found that at approximately
$R=10\arcsec$ (1.1 kpc) the stellar rotation changes its sense; the
maximum rotation velocity of the inner core, achieved at $R\approx 4\arcsec$,
was however rather low, of $\sim 15$ km/s. \citet{car97}
inspected high-resolution images obtained with the WFPC2/HST through two
filters, F555W and F814W, for the sample of elliptical galaxies with
kinematically decoupled cores.  In NGC~3608 they have not found 
any evidence for a circumnuclear stellar disk presence. Also they claimed
that the color $V-I$ is very homogeneous over the central parts of all
their sample galaxies, the kinematically decoupled cores being indistinctive
from the surrounding stellar bodies. In the particular case of
NGC~3608, however, we can suspect a break of $V-I$ by less that 0.1 mag at
$R\approx 4\arcsec$ from their Fig.~1{\it d}. Curiously, though either
\citet{car97} nor another investigators do not find any fine morphological 
substructure in NGC~3608, the global properties of the galaxy look unusual: 
despite its low luminosity, it is a slow rotator \citep{js88},
it is boxy \citep*{bendphot}, and it has a shallow core
profile \citep{car97} -- a classic set of properties
of a giant elliptical. General views of both galaxies are presented
by Figs.~\ref{maps3607} and \ref{maps3608} where three various fields 
of view are outlined: the large-scale view of the galaxies, the HST/PC-frame
field, and the very central regions which we have
observed by means of 2D spectroscopy.

The layout of the paper is as follows. We report our observations and
other data which we use in Section~2. 
In Section~3 two-dimensional velocity fields obtained
by means of 2D spectroscopy for the central parts of NGC~3607
and NGC~3608 are presented. The radial variations of the
stellar population properties are analysed in Section~4.
Section~5 gives discussion and our conclusions.

\section{Observations and data reduction}

The spectral data which we analyse in this work are obtained with
two different integral-field spectrographs. Integral-field spectroscopy
is a rather new approach which was firstly proposed by Prof. G. Courtes
some 20 years ago -- for a description of the instrumental idea see
e.g. \citet{betal95}. It allows to obtain simultaneously a set of
spectra in a wide spectral range from an extended area on the sky,
for example, from a central part of a galaxy.
In the spectrographs which data we use a 2D array of microlenses
provides a set of micropupils which are put onto the entry of
a spectrograph. After having reduced the full set of spectra
we obtain a list of the following characteristics for the individual spatial 
elements: fluxes in continuum and in emission lines,
line-of-sight velocities, both for stars and ionized gas, and
absorption-line equivalent widths which are usually expressed as
indices in the well-formulated Lick system \citep{woretal}.
This list can be transformed into two-dimensional maps of the
above mentioned characteristics for the central part of a galaxy
which is studied. Besides the panoramic view benefits, such an approach
gives an unique opportunity to overlay various 2D distributions
over each other without any difficulties with positioning.
In this work we use the data of two 2D spectrographs: the fiber-lens
Multi-Pupil Fiber Spectrograph (MPFS) at the 6m telescope of the Special
Astrophysical Observatory of the Russian Academy of Sciences (SAO RAS)
and the international Tiger-mode SAURON at the 4.2m William Herschel Telescope
at La Palma. SAURON is a private spectrograph, but some its raw data are
available after expiration for everybody from the Isaac
Newton Group Archive in the UK Astronomy Data Center.

The last variant of the MPFS became operational
in the prime focus of the 6m telescope in 1998
(http://www.sao.ru/hq/lsfvo/devices/mpfs/mpfs.html).
With respect to the previous variant, in the new MPFS
the field of view has been  increased and the common spectral range
has become larger due to using fibers. In 2001--2002 the fibers
 transmitted light from $16\times 15$ square
elements of the galaxy image to the slit of the spectrograph together
with additional 16 fibers that transmitted the sky background light
taken 4 arcminutes apart from the galaxy, so that the separate 
sky exposures were not necessary. The size of one spatial element was 
approximately $1\arcsec \times 1\arcsec$; a CCD TK $1024 \times 1024$ detector
was used before 2003. The reciprocal dispersion was 1.35~\AA\ per pixel, with
a spectral resolution of 4--5~\AA\ slightly varying over the field of view.
To calibrate the wavelength scale, we exposed separately a spectrum
of the hollow cathod lamp filled with helium, neon, and argon;
an internal accuracy of linearization was typically 0.25~\AA\ in the
green and 0.1~\AA\ in the red, and additionally we checked the accuracy
and an absence of systematic velocity shift by measuring strong emission
lines of the night sky [OI]$\lambda$5577 and [OI]$\lambda$6300.
We obtained the MPFS data in two spectral ranges,
the green one, 4300--5600~\AA, and the red one, 5900--7200~\AA.

The green-band spectra are used to calculate
the Lick indices H$\beta$, Mgb, Fe5270, and Fe5335 which are suitable
to determine metallicity, age, and Mg/Fe ratio of old stellar populations
\citep{worth94}. Following the prescriptions of \citet{worth94},
we have observed 15 stars from the list of \citet{woretal}
during four observational runs to calibrate the new MPFS index system 
onto the standard Lick one. After that we have calculated the
linear regression formulae to transform our index measurements into the
Lick system.  Below we give the regression formulae used by us
for the index calibration into the standard Lick system:\\
\noindent
H$\beta = (0.79 \pm 0.09) \mbox{H}\beta_{MPFS} +(0.18\pm 0.13)$ 
 
\noindent
Mgb $= (1.04 \pm 0.04) \mbox{Mgb}_{MPFS} -(0.26\pm 0.16)$ 
 
\noindent
Fe5270 $ = (0.98 \pm 0.05) \mbox{Fe5270}_{MPFS} -(0.32\pm 0.18)$ 
 
\noindent
Fe5335 $= (1.07 \pm 0.11) \mbox{Fe5335}_{MPFS} -(0.78\pm 0.40)$ 
 
\begin{figure}
\resizebox{\hsize}{!}{\includegraphics{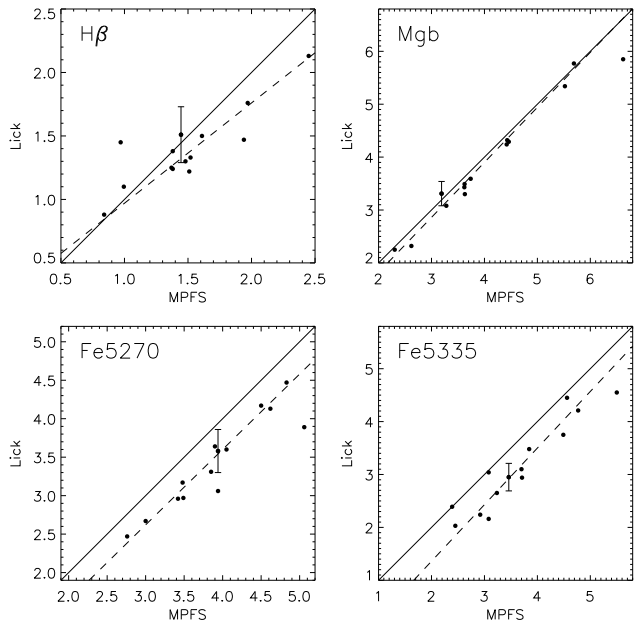}}
  \caption{The correlation of the instrumental absorption-line indices
obtained with the spectrograph MPFS and the standard Lick indices
for 15 stars from the list of Worthey et al. (1994). The thin
straight lines are bisectrices of the quadrants, the dashed lines are
linear regressions fitted to the dependencies. The mean index error
for the measurements of the stars by Worthey et al. (1994) are
indicated for a single star at every plot; our errors are lower by an
order.}
\label{mpfssys}
\end{figure}

\noindent
The comparison of star measurements and the linear calibration dependencies
are presented in Fig.~\ref{mpfssys}. The rms scatters of points near the linear
dependencies are less than 0.15~\AA\ for 3 indices except Fe5335 for which
it is 0.27~\AA, so the individual deviations of stars are within observational 
errors of \citet{woretal}.

We would like to stress that we do not degrade the spectral resolution
of our spectra. The Lick indices H$\beta$, Mgb, Fe5270, and Fe5335
are absorption-line equivalent widths by their definition, and equivalent
widths of the spectral lines do not depend on spectral resolution: they are integrals,
so by degrading the spectral resolution you make the lines shallower
but broader, and  integrals remain the same. The only restriction
is put by the wavelength intervals where the integrals are calculated.
The Lick index system is arranged to include
the full line width with the Lick spectral resolution, which is in average
$\sim 8$\AA, for the majority of galaxies. Both the MPFS and the SAURON 
have the better spectral resolution, so in our case the absorption lines
under consideration are completely within the wavelength intervals
prescribed by the Lick system, so we need no to make something with our
spectral resolution. This fact is confirmed by the close coincidence
of the instrumental index measurements of the standard stars with
the SAURON and the tabular index data from \citet{woretal} \citep{afsil02b}. 
However, small variations of the
continuum shape caused by imperfect spectra calibrations which are 
individual for every spectrograph may give small systematic deviations
of the instrumental indices from the Lick index system, and just 
this effect is corrected with the calibration formulae given above.
In the range of indices used, these corrections do not exceed
0.4~\AA\ for H$\beta$ and Fe5270, and 0.6~\AA\ for Fe5335;
for Mgb they are negligible. For the more detailed description
of the Lick index measurements with the MPFS and internal
accuracy estimates, see the recent
paper by \citet{silapj} where the study of the large sample of 
lenticular galaxies with the MPFS is presented.

To correct the index measurements for the stellar velocity dispersion
which is usually substantially non-zero in the centers of early-type
galaxies, we have fulfilled smoothing of the spectrum of the standard
star, HD~97907, by a set of running Gaussians of various widths; the derived
dependencies of index corrections on $\sigma _*$ were approximated by
polynomials of 4th order and applied to the measured index values
before their calibrations into the Lick system. Due to surface brightness
differences between the circumnuclear and more outer part of the bulges,
the accuracy of the measured indices falls from 0.1~\AA\ at the center
to more than 0.5~\AA\ at the edges of the multi-pupil frame. To support
a constant level of the index accuracy along the radii when analysing
the stellar population properties, we co-add the individual spectra
within circular rings centered onto a nucleus; the typical
statistical error of the azimuthally averaged indices is 
0.1~\AA --0.15~\AA\ \citep{silapj}.

The green-range 2D spectroscopic observations are also used to
cross-correlate galactic elementary spectra with a spectrum
of a template star, usually of G8III--K3III spectral type, to obtain
in such a way a line-of-sight velocity field for the stellar component
and a map of stellar velocity dispersion. The stars are defocused
during the observations to take into account the variations of
spectral resolution over the field of view. For NGC~3607 in particular,
we have calculated the kinematical maps with three templates
of different spectral types: HD 73665 (G8III), HD 73710 (G9III),
and the main component of STF 1947 (K0III). We have checked
that the kinematical maps obtained for NGC 3607 with three
different templates do not differ within the errors.  The cross-correlation peaks
are fitted by Gaussians; the benefits of this approach as opposed to the
more popular Fourier Correlation Quotient (FCQ) are thoroughly discussed
by \citet{bottema88}; we would like to stress that the cross-correlation method
is not so sensitive to template mismatch as FCQ or direct-fitting methods.
The red spectral range contains the strongest optical emission line for LINERS, 
[NII]$\lambda$6583,
so it is used to derive line-of-sight velocity fields for the ionized gas 
by calculating emission-line baricenter positions for [NII]$\lambda$6583.
The accuracy of elementary velocity measurements, both for stars and
ionized gas, is about 10 km/s.

The second integral-field spectrograph which data we use in this work is a new
instrument, the SAURON, operated at the 4.2m William Herschel 
Telescope (WHT) on La Palma -- for its detailed description see \citet{betal01}.
We have taken the data for NGC~3608
from the open ING Archive of the UK Astronomy Data Centre. Briefly,
the field of view of this instrument is $41\arcsec \times 33\arcsec$
with the spatial element size of $0\farcs 94 \times 0\farcs 94$.
The sky background taken 2 arcminutes from the center of the galaxy
is exposed simultaneously with the target. The fixed spectral range is
4800-5400~\AA, the reciprocal dispersion is 1.11~\AA\--1.21~\AA\
varying from the left to the right edge of the frame, and the
spectral resolution is about of 4~\AA. The comparison
spectrum is neon one, and the linearization is made by a polynomial
of the 2nd order with an accuracy of 0.07~\AA. The index system is
checked by using stars from the list of \citet{woretal} which have
been observed during the same observational run. The regressions fitting
the index system calibration of the February-1999 run when NGC~3608
has been observed are shown in our paper \citep{afsil02b}.
The relations between instrumental and standard-system indices
are very close to the $y=x$ relation so no corrections
are needed to calibrate them into the standard Lick system.
The stellar velocity dispersion effect has been corrected in the
same manner as for the MPFS data.
While to prepare the azimuthally averaged index profiles from
the MPFS data we co-added the spectra in the rings, to
prepare the SAURON azimuthally-averaged index data which have
higher signal-to-noise ratios,
we averaged the measured individual-element indices over the same rings so
the attached error bars are the errors of the means formally calculated -- 
they are all below 0.03~\AA.

The full list of the exposures made for NGC~3607 and
NGC~3608 with two 2D spectrographs is given in Table~2.
`BTA' means `Bolshoi Telescope Azimuthal'ny', it is an official name
of the 6m telescope. 

\begin{table*}
\caption[ ] {2D spectroscopy of the galaxies studied}
\begin{flushleft}
\begin{tabular}{lllllcc}
\hline\noalign{\smallskip}
Date & Galaxy & Exposure & Configuration & Field
& Spectral range & Seeing ($FWHM_*$) \\
\hline\noalign{\smallskip}
29 Apr 01 & NGC~3607 & 45 min & BTA/MPFS+CCD $1024 \times 1024$ &
$16\arcsec \times 15\arcsec $ & 4200-5600~\AA\ & $2\farcs 3$ \\
09 Mar 02 & NGC~3607 & 45 min & BTA/MPFS+CCD $1024 \times 1024 $ &
$16\arcsec \times 15\arcsec $ & 5800-7200~\AA\ & $2\farcs 8$ \\
09 Mar 02 & NGC~3608 &  60 min & BTA/MPFS+CCD $1024 \times 1024 $ &
$16\arcsec \times 15\arcsec $ & 5800-7200~\AA\ & $2\farcs 8$ \\
20 Feb 99 & NGC~3608 & 120 min & WHT/SAURON+CCD $2k\times 4k$ &
$33\arcsec\times 41\arcsec$ & 4800-5400~\AA\ & $1\farcs 4$ \\
\hline
\end{tabular}
\end{flushleft}
\end{table*}

To analyse the structure of the galaxies and to refine a kinematical
analysis, for both galaxies we have retrieved the WFPC2/HST image
data from the HST Archive. NGC~3607 was observed in the frame
of the program of A. Phillips (ID 5999) on nuclei of S0s, and
NGC~3608 was observed in the frame of the program of M. Franx on kinematically
decoupled nuclei (ID 5454). For NGC~3607 we have also used the
large-scale images taken from the database HYPERCAT/FITS Archive
(PI Ph. Prugniel). The details of the photometric observations are given
in Table~3.

\begin{table*}
\caption[ ] {Photometric data on NGC~3607 and NGC~3608}
\begin{flushleft}
\begin{tabular}{ccccccc}
\hline\noalign{\smallskip}
Date & Galaxy & Telescope & Filter & Exposure & Seeing & Scale \\
\hline\noalign{\smallskip}
7 May 1994 & NGC~3608 & WFPC2/HST & F555W & 500 s
& $0\farcs 1$ & $0\farcs 045$\\
7 May 1994 & NGC~3608 & WFPC2/HST & F814W & 230 s
& $0\farcs 1$ & $0\farcs 045$\\
6 Nov 1994 & NGC~3607 & WFPC2/HST & F547M & 260 s
& $0\farcs 1$ & $0\farcs 045$\\
27 Jan 1996 & NGC~3607 & WFPC2/HST & F555W & 160 s
& $0\farcs 1$ & $0\farcs 045$\\
27 Jan 1996 & NGC~3607 & WFPC2/HST & F814W & 160 s
& $0\farcs 1$ & $0\farcs 045$\\
6 Feb 1999 & NGC~3607 & 1.2-OHP & $B$ & 300 s
& $6\farcs 3$ & $0\farcs 686$\\
6 Feb 1999 & NGC~3607 & 1.2-OHP & $V$ & 120 s
& $6\farcs 0$ & $0\farcs 686$\\
6 Feb 1999 & NGC~3607 & 1.2-OHP & $R$ & 120 s
& $6\farcs 0$ & $0\farcs 686$\\
6 Feb 1999 & NGC~3607 & 1.2-OHP & $I$ & 120 s
& $5\farcs 7$ & $0\farcs 686$\\
\hline
\end{tabular}
\end{flushleft}
\end{table*}

All the data, spectral and photometric, except the data obtained with
the MPFS, have been reduced with the software produced by Dr. V.V. Vlasyuk
in the Special Astrophysical Observatory \citep{vlas}. Primary
reduction of the data obtained with the MPFS was done within IDL with
a software created by one of us (V.L.A.). The Lick indices were calculated
with our own FORTRAN program as well as by using the FORTRAN program of
Dr. A. Vazdekis which provides also the calculation of statistical errors
for the indices.

\section{Stellar and gaseous kinematics and the structure of the central
kiloparsecs}

\begin{figure}
 \resizebox{\hsize}{!}{\includegraphics{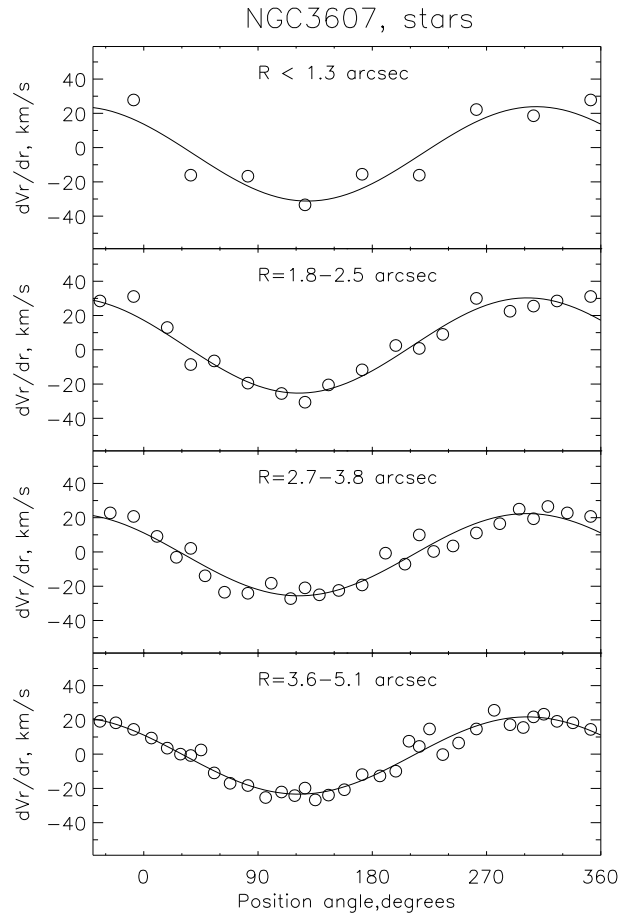}}
  \caption{The azimuthal dependencies of the central stellar line-of-sight
  velocity gradients taken in several radial bins for NGC~3607 according
  to the MPFS two-dimensional velocity measurements.}
 \label{cos3607}
\end{figure}

\begin{figure}
 \resizebox{\hsize}{!}{\includegraphics{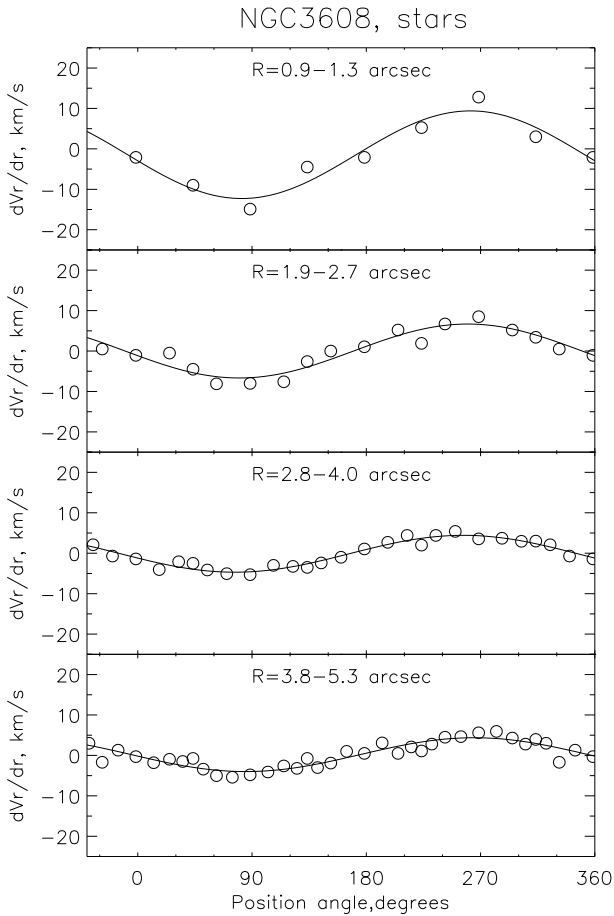}}
  \caption{The azimuthal dependencies of the central stellar line-of-sight
  velocity gradients taken in several radial bins for NGC~3608 according
  to the SAURON two-dimensional velocity measurements.}
 \label{cos3608}
\end{figure}

Since the integral-field spectroscopy provides us with two-dimensional
line-of-sight velocity fields, we are able now to analyse both
character of rotation and central structure of the galaxies.
If we have an axisymmetric mass distribution and mean rotation
on circular orbits around the symmetry axis,
the direction of maximum central
line-of-sight velocity gradient (we shall call it `kinematical major
axis') should coincide with the line of nodes  of the main symmetry
plane as well as the photometric major axis should do. In the case of 
a triaxial shape of potential the isovelocities  would tend to align 
with the principal axis of the ellipsoid \citep[][simulations of bars
and references therein]{vd97}. In a general case of triaxial potential
the kinematical and photometric major axes diverge,
showing turns with respect to the line of nodes in opposite senses
if the main axis of the triaxial potential is not strictly aligned
with the line of nodes \citep*{mbe92,mm2000}.
In the simplest case of thin-disk rotation we have a convenient
analytical expression for the azimuthal dependence of the central
line-of-sight velocity gradients within the area of solid-body rotation:\\

\noindent
$dv_r/dr = \omega \sin i \cos (PA - PA_0)$, \\

\noindent
where $\omega$ is the deprojected central angular rotation velocity,
$i$ is the inclination of the rotation plane, and $PA_0$ is the
orientation of the line of nodes  of the rotation plane.
In more complex three-dimensional cases the validity of the
formula above is not so evident; however the observed dependencies of
$dv_r/dr$ on $PA$ also look harmonic as one can see in
Figs.~\ref{cos3607} and \ref{cos3608}. Perhaps, it results from symmetry
properties that vertical and radial velocity components vanish after
integrating along the line of sight. At least it was shown that 
axisymmetric oblate spheroids
demonstrate zero line-of-sight velocity gradients along their
minor axes \citep{binney85} and the maximum rotation along their
major axes \citep{bacon85}. So by fitting azimuthal variations
of the central line-of-sight velocity gradients by a cosine
law, we can check the axisymmetry of the mass (stellar?)
distribution and to determine the orientation of the kinematical major
axis as its phase and the central angular rotation velocity
as its amplitude. It is our main tool for kinematical analysis.

\subsection{NGC~3607}

\begin{figure*}
\centering
  \includegraphics[width=17cm]{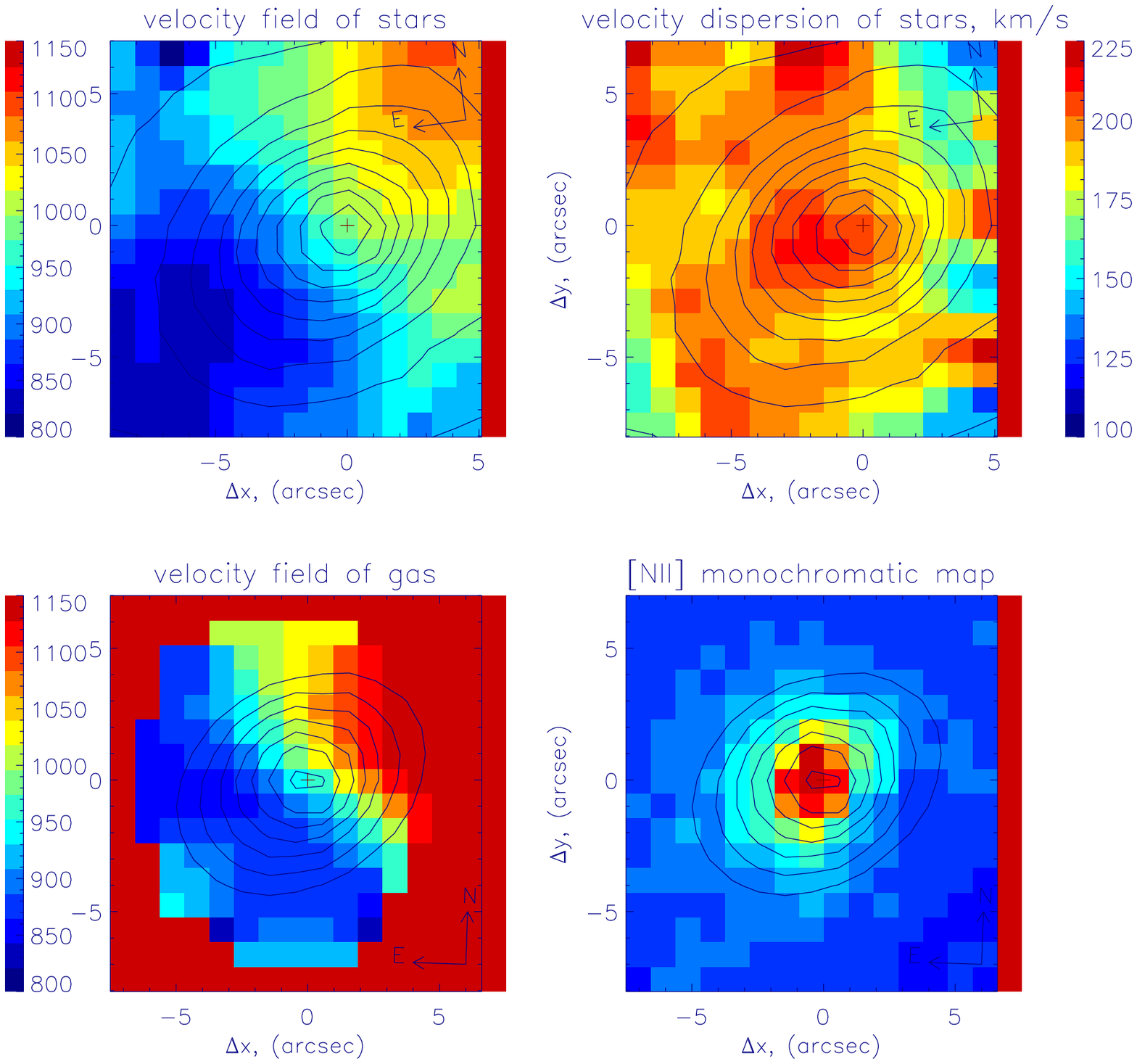}
  \caption{The line-of-sight velocity fields of the stellar component
  (top left) and of the ionized gas according to the measurements
  of [NII]$\lambda$6583 (bottom left),
  and the stellar velocity dispersion map (top right) and the
  [NII] emission line intensity distribution in arbitrary units (bottom right)
  in the central part of NGC~3607. The continuum, green 
 ($\lambda$5000~\AA) for  the stars and red ($\lambda$6400~\AA) for the gas, 
 is presented by isophotes.}
 \label{kin3607}
\end{figure*}

Figure~\ref{kin3607} presents mostly kinematical maps for the central
part of NGC~3607.
Both stars and ionized gas demonstrate regular, almost rigid-body rotation
with their kinematical major axes close to the photometric major axis.
The gas rotation is twice faster than that
of the stars: the central angular rotation velocity is
$\omega _{gas} \sin i \approx 50$ km/s/arcsec whereas
$\omega _{stars} \sin i \approx 28$ km/s/arcsec. Let us to note that if
the inclination given by LEDA for the symmetry galactic plane and derived from
the large-scale disk axis ratio, $34\degr$ (Table~1), is correct, 
it means that the central gas rotates at a speed of
800 km/s/kpc. However, below we should recognize that it is rather difficult
to determine precise orientation(s) of the rotation plane(s) in the center
of NGC~3607. The stellar velocity dispersion has a peak near
the nucleus.  In our data this stellar velocity dispersion maximum
area is not exactly centered onto the nucleus that may be a real effect
due to the asymmetric dust distribution in the center of the galaxy; this
maximum area seems to be extended in the direction of the isophote minor axis.

\begin{figure}
 \resizebox{\hsize}{!}{\includegraphics{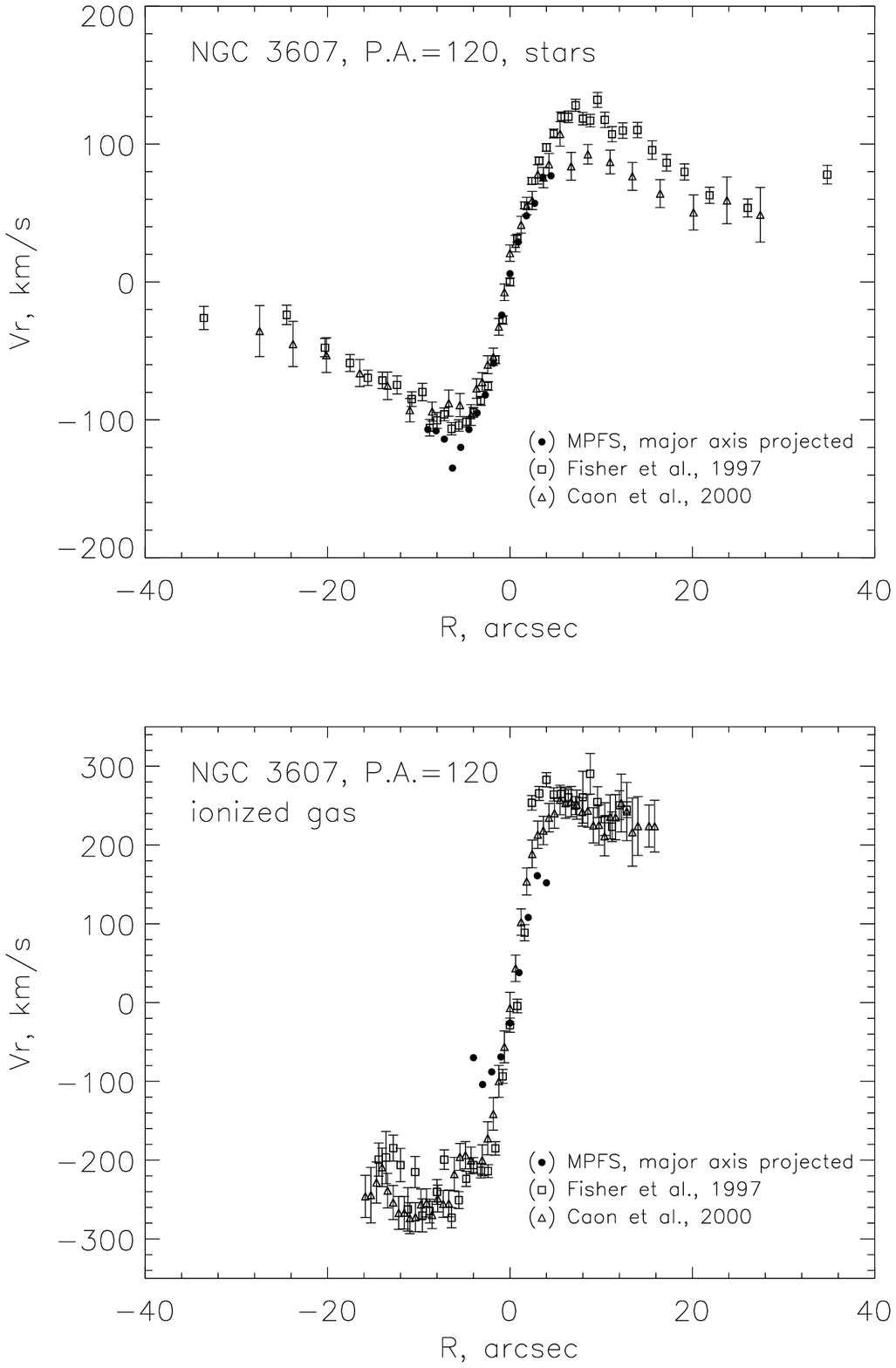}}
 \caption{The comparison of the line-of-sight velocity profiles
 simulated along the major axis by using our 2D velocity fields
 for the stars and ionized gas in NGC 3607 with the literature data
 of Fisher (1997) and of Caon et al. (2000).
 The slit width used in the simulations is 2\arcsec.
 }
\label{cuts3607}
\end{figure}

\begin{figure}
 \resizebox{\hsize}{!}{\includegraphics{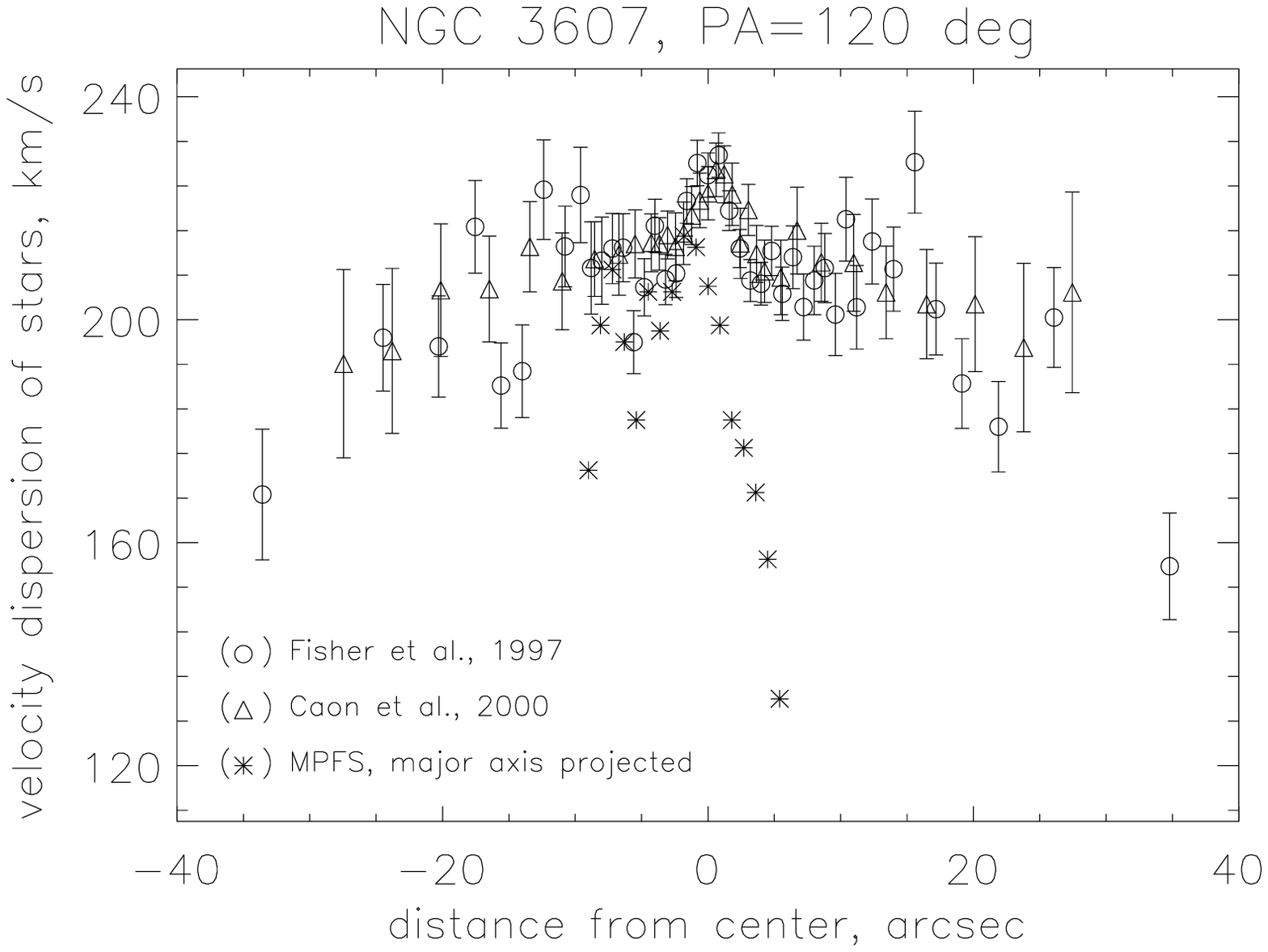}}
 \caption{The comparison of the stellar velocity dispersion profiles
 simulated along the major axis by using our 2D velocity fields
 for the stars and ionized gas in NGC 3607 with the literature data
 of Fisher (1997) and of Caon et al. (2000).
  The slit width used in the simulations is 2\arcsec.
 }
\label{cutw3607}
\end{figure}

To check our 2D mapping of the line-of-sight velocities, we have simulated
one-dimensional cross-sections of the velocity fields of Fig.~\ref{kin3607}:
a band aligned with the major axis has been overposed onto the maps
and the velocity and dispersion velocity values inside this band have been
derived and averaged over the $1\arcsec \times 2\arcsec$ elements.
We have compared the simulated LOS velocity and stellar velocity
dispersion profiles with the long-slit data
along the major axis of NGC~3607 which we have found in the
literature (Fig.~\ref{cuts3607} and \ref{cutw3607}). One can see that the 
agreement of the velocity profiles is rather good.  However, the 
literature data do not support the
shift of the velocity dispersion peak with respect to the photometric
center. Also, perhaps the slope of the ionized gas
velocity profile in our data may be slightly underestimated due to moderate
seeing quality of the observations of March 2002. The maximum gas rotation
velocity, $\sim 250$ km/s at $R=5\arcsec -10\arcsec$ which may be transformed
into 450 km/s under $i=34\degr$, seems to be enormous; but this result is
confirmed by recent CO observations of NGC~3607 according
to which $2.3 \cdot 10^8 M_{\odot}$ of the molecular gas within
$R=5\arcsec$ rotates at a speed of 250 km/s \citep{co_s0}.
Interestingly, in the integrated
molecular-line profile only a receding horn is well seen -- the asymmetry
is quite similar to that of the Fig.~\ref{kin3607} (the lower left).
The stellar velocity profile of Fig.~\ref{cuts3607} reveals a prominent
rotation velocity maximum at $R\approx 6\arcsec$
after which the rotation velocity drops by a factor of $\sim 3$ toward
$R=20\arcsec - 25\arcsec$. Such behaviour may be treated as a kinematically
distinct core in the center of NGC~3607;
let us to note that the maximum rotation radius coincides exactly with
the border of the magnesium-enhanced structure (see the next Section).

\begin{figure}
 \resizebox{\hsize}{!}{\includegraphics{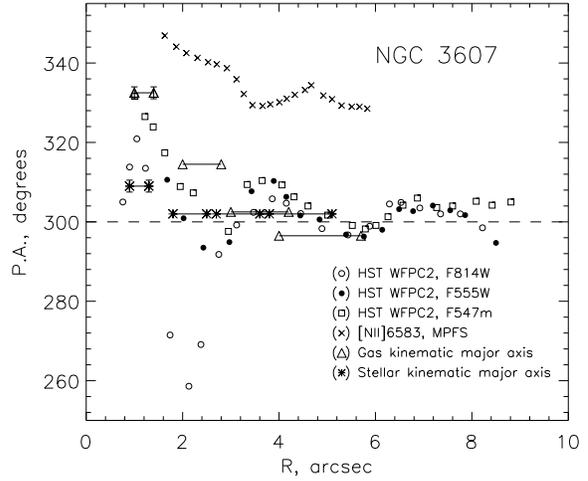}}
 \caption{Isophote major axis position angle compared to the orientations of
    the kinematical major axes (see the text) for the stars and ionized gas
    in the center of NGC~3607. The errors of the kinematical major axes
    determination are estimated as 1--1.5 deg. The line of nodes determined
    from the outermost disk isophote orientation, $PA=300\degr$, is traced
    by a dashed line.}
 \label{isocomp1}
\end{figure}

\begin{figure}
 \resizebox{\hsize}{!}{\includegraphics{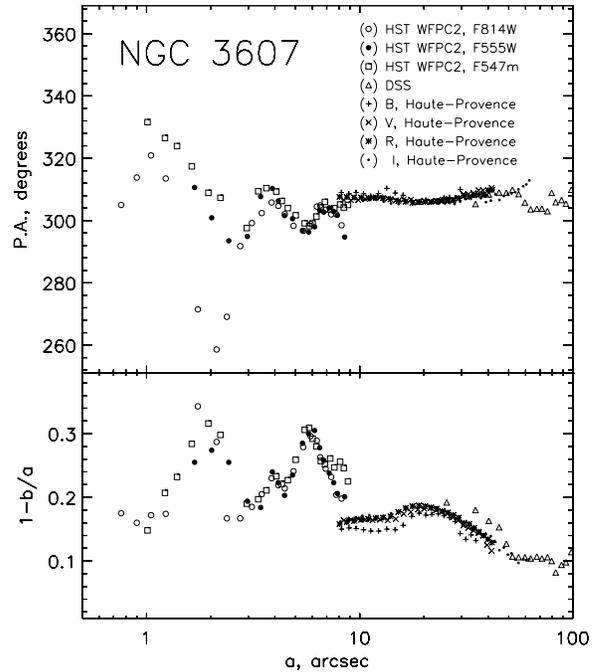}}
 \caption{Isophote characteristics in the full radius range
    of NGC~3607.}
 \label{iso3607}
\end{figure}

The enormous rotation velocity, which we have measured in the center
of NGC~3607 and which implies a mass concentration of more
than $10^{10} M_{\odot}$ within $R\approx 0.5$ kpc, provokes some doubts
about a circular (axisymmetric) character of rotation. We can check it by
comparing kinematical and photometric major axis orientations; it is
done in Fig.~\ref{isocomp1}. Our first impression from inspecting
Fig.~\ref{kin3607} is that the kinematical and photometric major axes
coincide; the Fig.~\ref{isocomp1} confirms this impression by quantifying it. 
At $R>3\arcsec$, or more exactly in the radius range of
$R=3\arcsec - 6\arcsec$,  within the decoupled core area, 
where the rotation is still almost solid-body, the kinematical
major axes of the ionized gas and of the stars, as well as the photometric
major axes of the continuum brightness distribution are all close to the
global line of nodes of the galaxy, $PA_0=300\degr (120\degr)$. This fact
agrees with an axisymmetric character of rotation with the spin
orthogonal to the main symmetry plane of the galaxy -- a classic
configuration corresponding to the SA0-type of NGC~3607. But
inside $R=2\arcsec$ ALL THE AXES -- both the kinematical ones and the
photometric one --  seem to turn toward larger
$PAs$, the turn of the gas kinematical major axis being the most prominent.
We have estimated an orientation of the innermost isophotes of
$\mbox{H}\alpha +$[NII] emission distribution in Fig.~1f of
\citet{shields} -- it is $PA\approx 325\degr$, very close to our measurements
of the [NII] emission-line brightness distribution elongation at $R>3\arcsec$. 
The gray-scaled emission brightness map provided by \citet{macetal}) confirms 
also the turn of the innermost emission isophotes toward larger $PAs$. 
The concordance of the photometric major axis provided by the HST data 
and the kinematical major axis of the ionized gas near 
$PA\approx 325\degr -330\degr$ at $R\approx 1\arcsec - 2\arcsec$ implies  
axisymmetry.  But the rotation  plane of the innermost, $R<2\arcsec$, part of the 
galaxy,  and so  that of circumnuclear disk probably,  appears to be inclined 
to the main symmetry plane of NGC~3607, while the more outer gas rotates 
in the global galactic plane together with the stars. The latter fact rejects 
an external origin of the decoupled central spin; there must be own intrinsic
peculiarities of NGC~3607 to provide such kinematical appearance.
Figure~\ref{iso3607} shows radial variations of the isophote characteristics
in NGC~3607 at larger scales than Fig.~\ref{isocomp1} does.
The isophote major axis position angle remains almost constant all along 
the radius. But the
ellipticity behaviour is indeed strange: it reaches a main maximum
of 0.3 at the border of the decoupled core, at $R=6\arcsec$, then it reaches
a secondary maximum of 0.2 at $R\approx 20\arcsec$, and after that it
falls to $1-b/a\approx 0.1$ at $R\approx 60\arcsec$. The formal inclination
of the global galactic disk, $i=34\degr$ (LEDA), corresponds only to
$1-b/a=0.17$ by assuming infinitely thin disk. Let us also to note that
the radius of the minimal ellipticity, $R\approx 60\arcsec$, coincides
with the boundary between the bulge and the global disk of NGC~3607:
according to \citet*{bag}, in NGC~3607 the bulge dominates
at $R< 60\arcsec$ and the exponential disk -- at larger radii. The ellipticity
behaviour at $R \geq 20\arcsec$ is roughly consistent in all filters,
so it is not an effect of dust. We must conclude that the bulge of
NGC~3607 cannot be an axisymmetric (oblate) spheroid  if to require
its symmetry plane to coincide with the disk plane, otherwise
its visible ellipticity should be smaller than that of the more thin
disk. It has to be triaxial at a scale of a few kiloparsec.

\subsection{NGC~3608}

\begin{figure}
  \resizebox{\hsize}{!}{\includegraphics{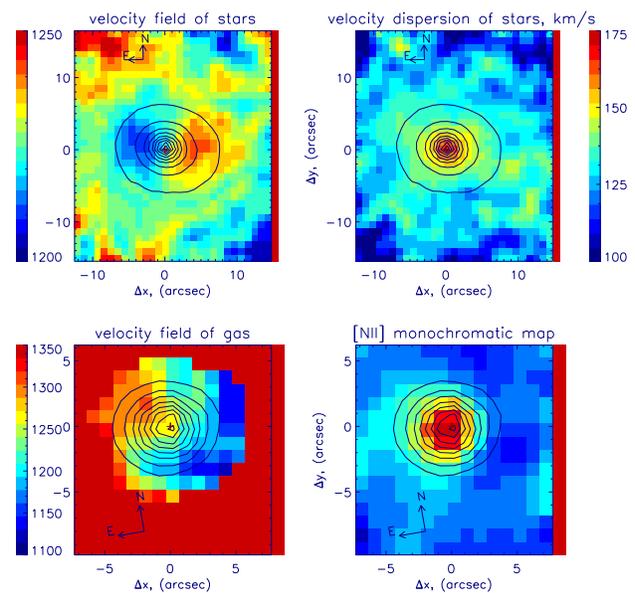}}
  \caption{The line-of-sight velocity fields of the stellar component
  (top left) and of the ionized gas according to the measurements
  of [NII]$\lambda$6583 (bottom left), and the stellar velocity dispersion
map (top right) and the [NII] emission line intensity distribution (bottom right)
  in the central part of NGC~3608. The continuum, green ($\lambda$5000~\AA)
  for the stars and red ($\lambda$6400~\AA) for the gas, is presented by 
  isophotes.}
 \label{kin3608}
\end{figure}

\begin{figure}
 \resizebox{\hsize}{!}{\includegraphics{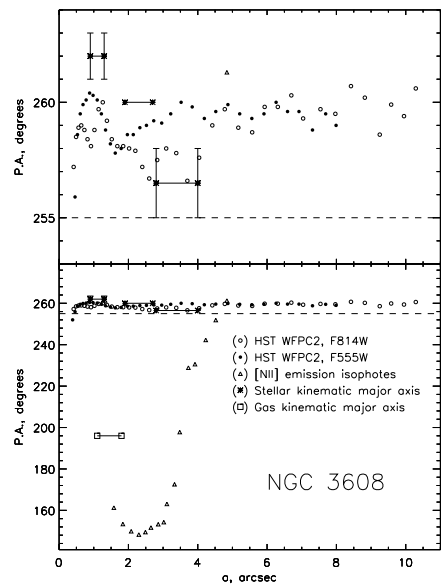}}
 \caption{Isophote major axis position angles compared to the orientations of
    the kinematical major axes (see the text) for the stars and ionized gas
    in the center of NGC~3608. The line of nodes determined from the
    outermost isophote orientation, $PA=255\degr$, is traced by
    a dashed line. The data for the stars are zoomed separately at the top
    plot to demonstrate the kinematic major-axis turn.
    The errors of the kinematical major axes
    determination are estimated as 1--1.5 deg.
    }
 \label{isocomp2}
\end{figure}

To analyse the stellar kinematics of the central part of NGC~3608 we
have used the SAURON data reduced by us with our tools. 
Recently the stellar kinematical maps for NGC~3608 have been  
published by \citet{sau3} together with the maps
of LOSVD higher moments $h_3$ and $h_4$; the SAURON team has applied 
its own method to analyze these data which is quite sophisticated.
The quantitative agreement of the LOS velocity fields obtained by us and by 
\citet{sau3} is perfect; our stellar velocity dispersion measurements
being qualitatively similar are systematically lower by about
of 40 km/s.  In the further discussion we use mainly the morphology
of the stellar velocity dispersion distribution, so the precise absolute 
values of $\sigma _*$ are not so important in this particular work.

In Fig.~\ref{kin3608} one can see the comparison
of the LOS velocity fields for the stars and for the ionized gas.
It reveals indeed a central counterrotating stellar
subsystem with the kinematical major axis close to the major axis of
the inner isophotes -- it is the kinematically decoupled core reported
by \citet{js88}. The ionized-gas velocity field
(Fig.~\ref{kin3608}, bottom) reveals a strong twist of isovelocites
over the whole measured velocity field. The amplitude of the
circumnuclear azimuthal variations of the gas line-of-sight velocity gradient
is $73 \pm 15$ km/s/arcsec -- even higher than in NGC~3607.
The phase of the cosine curve fitting the
azimuthal variations of the central gas line-of-sight velocity gradients
within $R=2\arcsec$ is shifted by $\sim 70\degr$ with respect to
the line of nodes of the stellar rotation.
As the nucleus of NGC~3608 is inactive and radio-quiet,
we would disregard a hypothesis of outflow and would claim
rather a kind of strongly inclined gaseous ring
with a radius of $\sim 200$ pc. In Fig.~\ref{isocomp2} 
we compare the orientations of
the kinematical and photometric major axes, for the stars and for the
ionized-gas emission [NII]$\lambda$6583. The kinematical and
photometric major axes of the stellar component coincide that implies
an axisymmetric rotation.  However both major axes begin to
deviate from the $PA$ of outer isophotes, $PA_0=255\degr$, in a systematic 
way when approaching the very center.
Meantime, the major axes of the [NII] emission brightness
distribution and of the ionized-gas velocity field 
are turned strongly with respect to the stellar
configuration at $R<4\arcsec$, or within the kinematically decoupled area.
The spatial resolution of our red-band MPFS exposures is rather moderate,
so the measurements inside $R=2\farcs 5$ are strongly affected by the
seeing effect. From the published HST data analysis
\citep{hstjapan} we know that in the center of NGC~3608 the dust ring with the radius
of $0\farcs 5$ is roughly aligned with the outer-isophote orientation,
$PA_{dust}\approx 80\degr$ (or $260\degr$) (see also our Fig.~\ref{maps3608},
right plot). Since  the dust and the gas are related,
the orientation of the circumnuclear gas plane must also be close to the
line of nodes, so one may extrapolate our measurements back
to the line of nodes inside $R\approx 1\arcsec$. Strong twist of both
the photometric and kinematical major axes for the gas component
at $R=1\arcsec - 4\arcsec$,
though being obviously smoothed by the seeing effect, implies neverthless
an existence of a kind of polar gas rotation in the above-mentioned
radius range.

\begin{figure}
 \resizebox{\hsize}{!}{\includegraphics{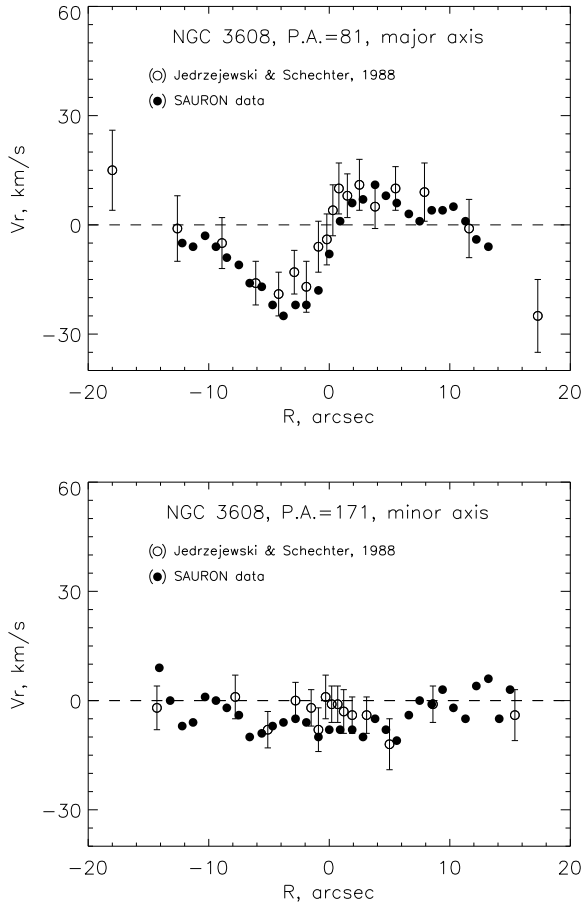}}
 \caption{The comparison of the line-of-sight velocity profiles
 simulated along the major and minor axes by using the SAURON 2D 
LOS velocity  field for the stars in the center of NGC 3608 with the literature data
 of Jedrzejewski and Schechter (1988). 
 The slit width used in the simulations is 2\arcsec.
 }
\label{cuts3608}
\end{figure}

\begin{figure}
 \resizebox{\hsize}{!}{\includegraphics{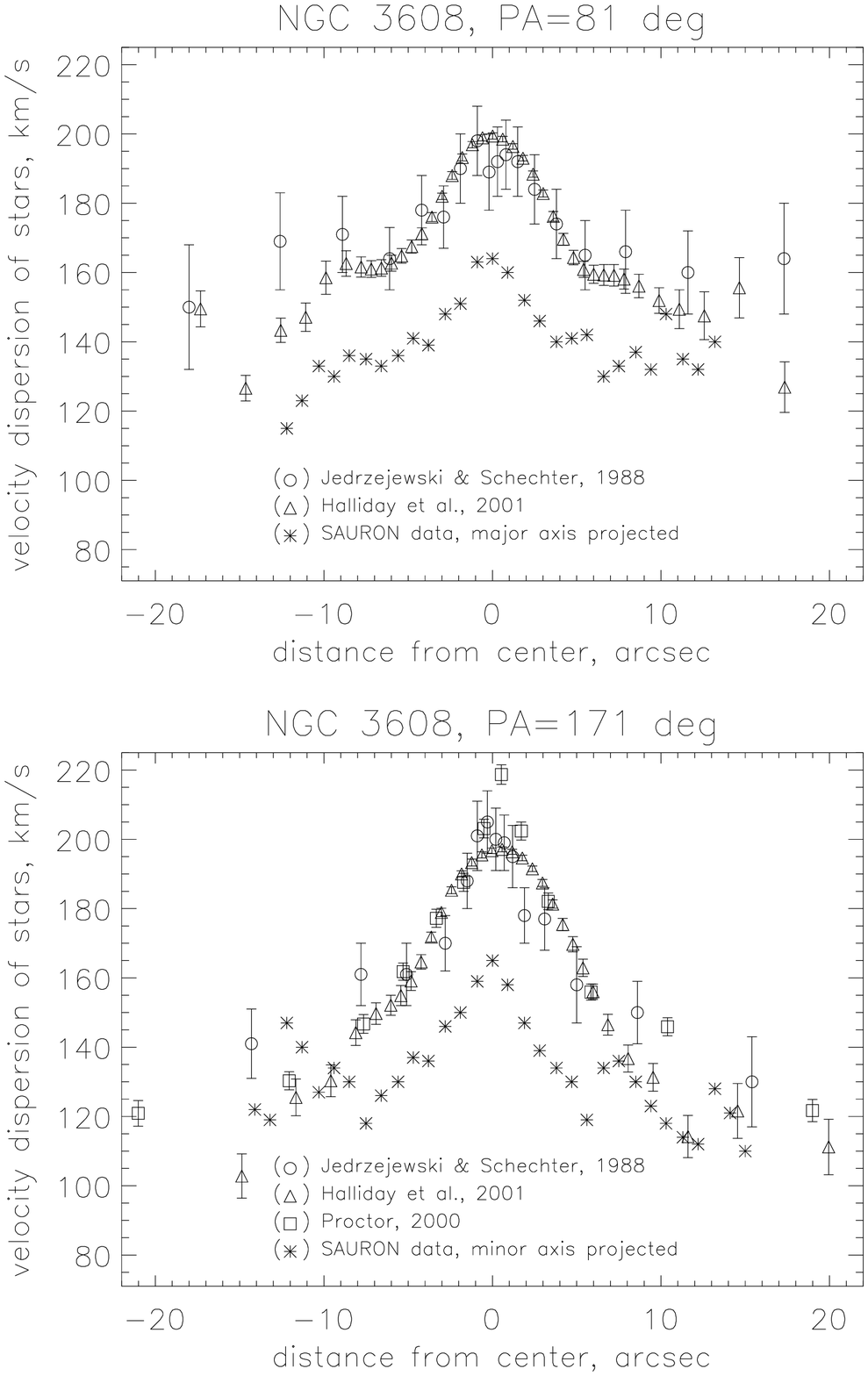}}
 \caption{The comparison of the stellar velocity dispersion profiles
 simulated along the major and minor axes by using the SAURON 2D velocity 
 dispersion field for the stars in the center of NGC 3608 with the literature 
 data of Jedrzejewski and Schechter (1988), Halliday et al. (2001),
 and Proctor (2002)
  The slit width used in the simulations is 2\arcsec.
 }
\label{cutw3608}
\end{figure}

Figures~\ref{cuts3608} and \ref{cutw3608} demonstrate simulated major- 
and minor-axis cross-sections of the stellar LOS velocity and velocity
dispersion fields in the center of NGC~3608 in comparison with published 
long-slit data. The agreement of our cuts with the literature long-slit data
is satisfactory though the systematic shift of our stellar velocity dispersion
measurements by $\sim -30 - -40$ km/s can be also seen.
We would note that systematic differences of a few dozen kilometre per
second are still common for stellar velocity dispersion measurements.
For example, in the recent study of 48 early-type galaxies by \citet{sau3}
the individual differences between the measured $r_e/10$-aperture
stellar velocity dispersions and the literature data reach 40 km/s (see
their Fig.~3). Partly this uncertainty may be due to using different methods
to derive kinematical information from the absorption-line spectra:
various methods are sensitive to signal-to-noise ratio and to template
mismatch by various degree. However \citet{bhsdisp} have shown that 
even using the same method over the data of similar quality does not save
the matter: the stellar velocity dispersion estimates made over different
spectral ranges may differ by 20--30 km/s. So we would not refer to the
exact absolute values of $\sigma _*$, but rather would analyse their
relative variations over the field of view.
 
Besides the overall check of the quality of our velocity measurements,
Figs.~\ref{cuts3608} and \ref{cutw3608} confirm once more that
the edge of the kinematically decoupled stellar core
is located at $R=4\arcsec$ and, as we shall see in the next Section, 
coincides roughly with the border of the magnesium-decoupled core. 
But the kinematical analysis
does not leave a possibility to treat the decoupled core as a compact
stellar disk. The  low $v/\sigma$ ratio, $\sim 0.1$, and a pattern of the stellar
velocity dispersion distribution -- an extended maximum clearly seen
in the stellar velocity dispersion map (Fig.~\ref{kin3608}, top right)
-- allow to suggest a triaxiality of the decoupled core. In such
configuration the polar rotation of the ionized gas becomes explicable:
we often meet circumnuclear polar gaseous disks in galaxies with
triaxial bulges or within bars -- e. g. in NGC~2841
\citep*{silvb97,we99}, in NGC~6340 \citep{me2000}, in NGC~7280
\citep{we7280}, in NGC~4548 \citep{me2002}, etc.

\section{Chemically decoupled cores in NGC~3607 and NGC~3608}

\subsection{NGC~3607}

\begin{figure*}
\centering
  \includegraphics[width=17cm]{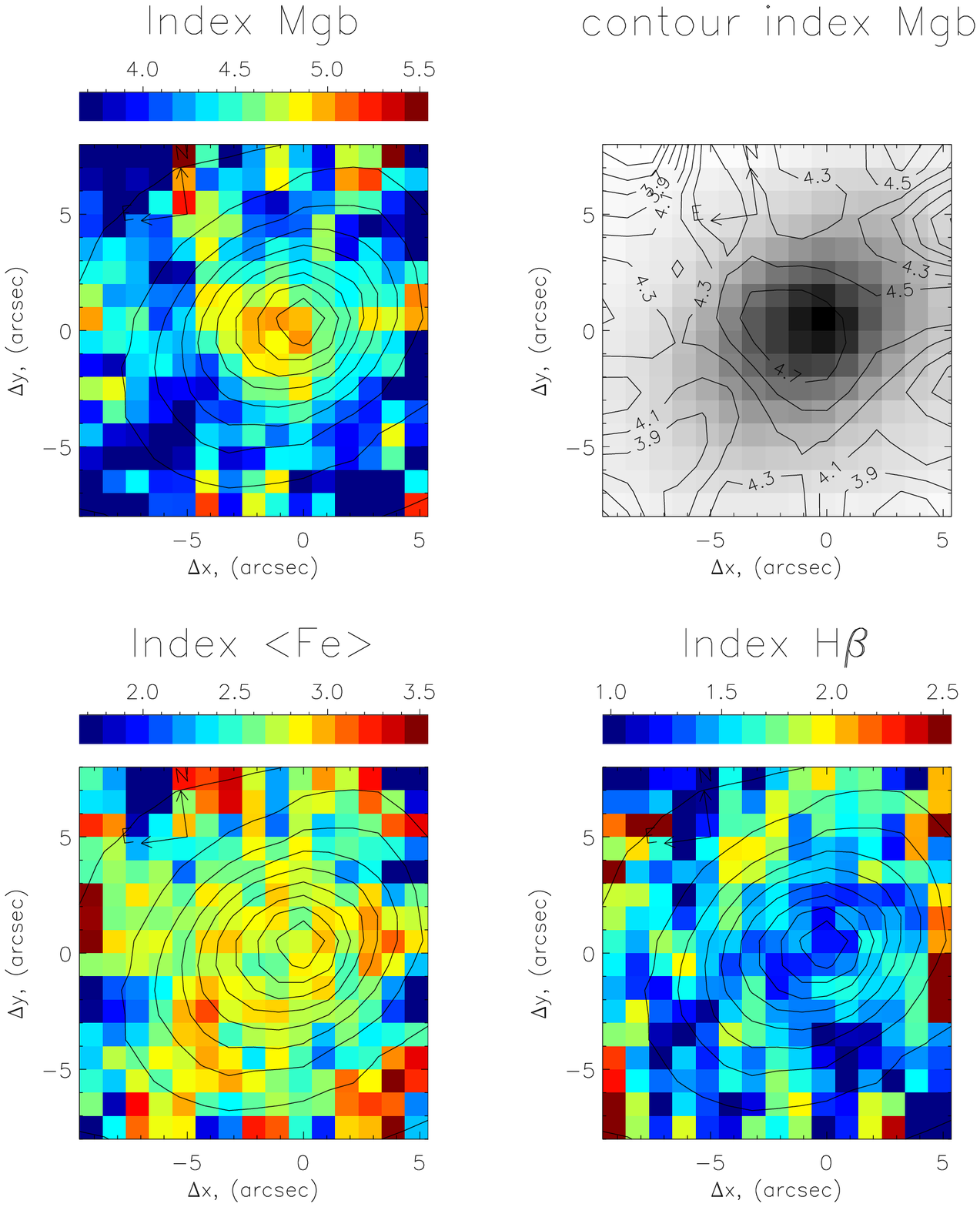}
  \caption{The MPFS index maps for NGC 3607;
  $<\mbox{Fe}> \equiv$(Fe5270+Fe5335)/2. The green
  ($\lambda$5000~\AA ) continuum is overlaid by isophotes in
  three coloured plots. At the right top the Mgb index distribution
  smoothed strongly with the 2D Gaussian of $FWHM=3\farcs 5$ is plotted
  by isolines to show an orientation of the chemically decoupled
  structure; here the green continuum is gray-scaled}
  \label{imap3607}
\end{figure*}

Figure~\ref{imap3607} presents Lick index maps for the central part of
NGC~3607 constructed from the MPFS 2D spectral data.
The Mgb map reveals a certain presence
of chemically decoupled core; this finding is not very unexpected because
NGC~3607 has been earlier listed by us as a good candidate for
possessing a chemically distinct nucleus \citep{sil94}.
Interestingly, the
$\langle \mbox{Fe} \rangle \equiv$(Fe5270+Fe5335)/2 and H$\beta$ absorption
line index maps look rather homogeneous and do not demonstrate any sharp
features. Perhaps, one can note a very shallow unresolved
$\langle \mbox{Fe} \rangle$ minimum near the photometric center
of the galaxy and similarly shallow extended H$\beta$ minimum aligned
with the major axis of the isophotes; the latter feature may be associated
to the circumnuclear ionized-gas disk mentioned in the Introduction
and is probably resulted from the emission contamination of the H$\beta$
absorption line. Interestingly, all the central peculiar features
of the index distributions are centered slightly below and to the left
from the photometric center showing the same shift as the stellar
velocity dispersion. However, both
$\langle \mbox{Fe} \rangle \equiv$(Fe5270+Fe5335)/2 and H$\beta$ 
absorption-line index minima are marginal, and
the Mgb peak is much more prominent than the other
details. It seems to be resolved; moreover, it seems to be elongated
along the minor axis of the isophotes. To derive the orientation of
the magnesium-enhanced core, we have smoothed strongly the Mgb map
and have drawn isolines of the smoothed Mgb distribution (Fig.~\ref{imap3607},
top right). Indeed, the magnesium-enhanced core is elongated in
$PA\approx 46\degr$ though the Mgb isolines do not look as quite symmetric
ovals. The orientation of the magnesium-enhanced area hints that we see
some stellar substructure with its alignment
along the minor axis of the isophotes -- some kind of minibar or so. But
in fact, the Mgb-index distribution morphology alone is not enough to
make a conclusion about the nature of the magnesium-enhanced core.
The visible elongation of the magnesium-enhanced core may be a result of
a magnesium depression along the major axis, similar to that
of the H$\beta$ index, which may be caused by
a very young stellar disk aligned with the major axis. To make
a definite conclusion, a further analysis of the stellar population
age in the area of magnesium-enhanced core is needed.

\begin{figure}
 \resizebox{\hsize}{!}{\includegraphics{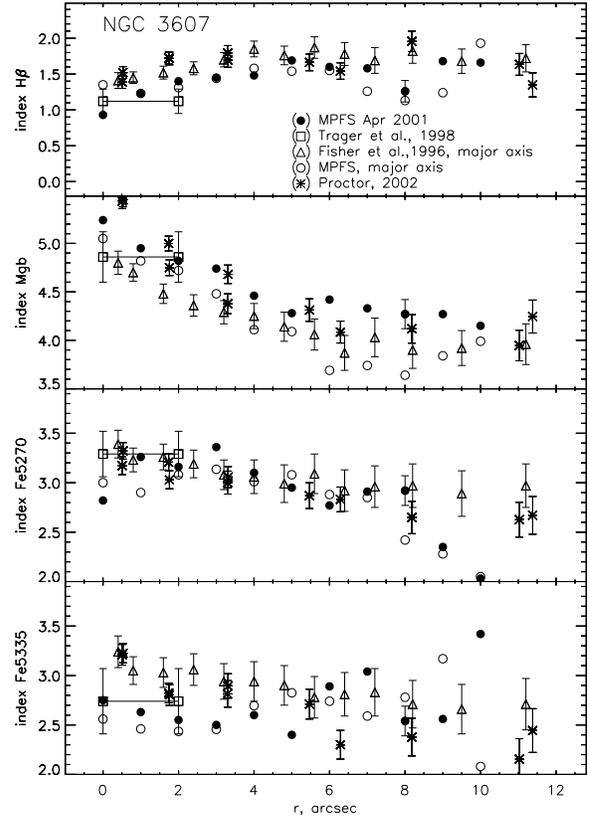}}
 \caption{The azimuthally-averaged index profiles for NGC~3607
 in comparison with the literature data. The minor-axis long-slit
 cross-section by Proctor (2002) is plotted. The long-slit data taken along
 the major axis by Fisher et al. (1996) are also compared to the
 major-axis index profiles simulated from our 2D index maps
 with the digital slit width of 2\arcsec. The central $2arcsec\ \times
 4\arcsec$ aperture measurements by Trager et al. (1998) are plotted
 as connected squares at $R=0\arcsec -2\arcsec$.}
 \label{indcomp1}
\end{figure}

To quantify peculiarities of the absorption-line index distributions
in the center of NGC~3607, we have calculated
azimuthally averaged radial profiles of the indices and have compared
them to the `zero-dimensional' and one-dimensional spectral data
published earlier by \citet{prdiss}, \citet{trager}, and
\citet*{fish96} (Fig.~\ref{indcomp1}). Though the long-slit data were taken  
along the major or minor axis of NGC~3607 and so were not
obliged to agree exactly with our azimuthally averaged measurements,
they confirm the overall shape of the index radial dependencies:
a rather flat iron profile, an H$\beta$ profile with a shallow minimum
in the center, and a magnesium-distinct core with the radius of 
$5\arcsec - 6\arcsec$. A moderate, 
by 0.2~\AA--0.4~\AA, systematic difference of our H$\beta$ and Mgb profiles 
with the data of \citet{fish96} can also be noted. To check if these shifts 
may be due to a lack of azimuthal symmetry, we have simulated 
one-dimensional major-axis cross-sections of our index maps and have 
also plotted the simulated results in Fig.~\ref{indcomp1}.
Indeed, the simulated major-axis Mgb profile has a much better
agreement with the Fisher's et al. data than the azimuthally
averaged measurements. As for the H$\beta$ index profiles, the
simulated major-axis profile and the azimuthally averaged one have
coincided almost perfectly, implying an orientation of the ionized-gas
disk far from edge-on.  We must conclude that the systematic shift
of $\sim$0.2~\AA\ between our measurements of the H$\beta$ index
and those of \citet{prdiss} and
\citet{fish96} exists really. Meantime, the centered aperture measurements
by \citet{trager} which  are a Lick-system etalon confirm rather our calibration.

\begin{figure}
 \resizebox{\hsize}{!}{\includegraphics{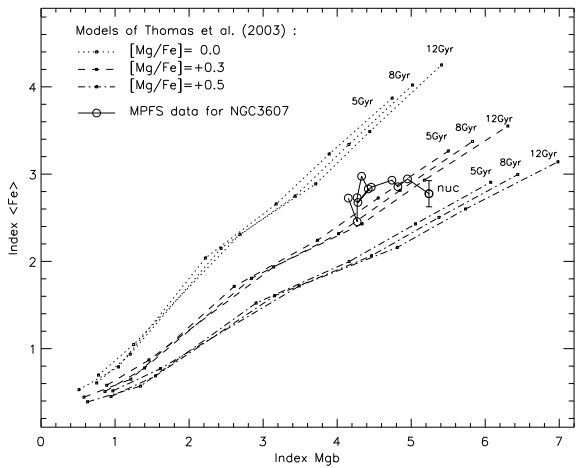}}
 \caption{`Index-index' $<\mbox{Fe}>$ vs Mgb diagnostic diagram
 for the azimuthally averaged  Lick indices in the center of NGC~3607
 taken along the radius with the step of 1\arcsec\ (open circles connected
 by a thin line, the nucleus being marked as `nuc'). The errors
 of the azimuthally averaged MPFS indices are about 0.1~\AA.
 The models of Thomas et al. (2003) for [Mg/Fe]=0.0, +0.3, and +0.5
 are plotted as a reference frame; the small signs connected by
 pointed, dashed, and point-dashed lines represent stellar population 
 models of equal ages; the metallicities for the models are
 +0.67, +0.35, 0.0, -0.33, -1.35, and -2.25 if one takes the signs from
 top to bottom}
 \label{mgfe3607}
\end{figure}

Due to progress in evolutionary population synthesis during last years,
we can now estimate mean stellar population characteristics by confronting
different absorption-line indices to each other. Figure~\ref{mgfe3607}
presents an `iron-vs-magnesium' diagram where we compare our azimuthally 
averaged MPFS data for the central part of NGC~3607 with the models
of \citet*{thomod} which are calculated for several values of 
magnesium-to-iron ratio. The loci of the models of various [Mg/Fe] are well
separated on the diagram, so from inspecting Fig.~\ref{mgfe3607} we can
conclude that the magnesium-to-iron ratio in the center
of NGC~3607 is certainly above the solar one. But whereas in the 
circumnuclear region the [Mg/Fe] is between zero and $+0.3$, 
perhaps, closer to $+0.2$, the unresolved nucleus
is outstanding with its [Mg/Fe]$\approx +0.4$. We must note that, as
Fig.~\ref{indcomp1} demonstrates, the previous Lick index measurements for
NGC~3607 do not confirm this jump of [Mg/Fe] at $R=0\arcsec$ and
imply rather constant iron-index behaviour along the radius. Since our
estimates of the nuclear indices are confined to a single spatial element,
we cannot insist that the low Fe5270 value in the unresolved nucleus is not
a random artifact. So over all the center of NGC~3607
we assume [Mg/Fe] to be between $+0.2$ and $+0.4$.

\begin{figure}
 \resizebox{\hsize}{!}{\includegraphics{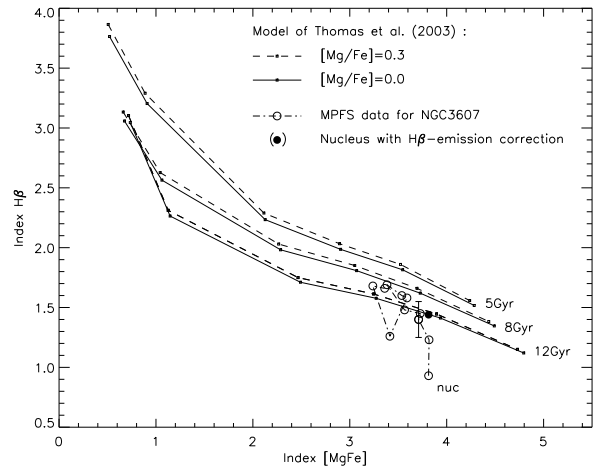}}
 \caption{`Index-index' H$\beta$ vs [MgFe],
  [MgFe]$\equiv (\mbox{Mgb} <\mbox{Fe}>)^{1/2}$, age-diagnostic diagram
 for the azimuthally averaged  Lick indices in the center of NGC~3607
 taken along the radius with the step of 1\arcsec\ (open circles connected
 by a dot-dashed line, the nucleus being marked by `nuc'). The errors
 of the azimuthally averaged MPFS indices are about 0.1~\AA.
 The models of Thomas et al. (2003) for [Mg/Fe]=0.0 and +0.3
 are plotted as a reference frame; the small signs connected by solid and
 dashed lines represent stellar population models of equal ages;
 the metallicities for the models are
 +0.67, +0.35, 0.0, -0.33, -1.35, and -2.25, if one takes the signs from
 the right to the left.}
 \label{age3607}
\end{figure}

Several investigators, e.g. \citet{gonzdis} or \citet{terlfor}, 
noted that in order to overcome an effect of 
the non-solar magnesium-to-iron ratio when determining a mean age 
of stellar population, one must use a combined index
[MgFe]$\equiv (\mbox{Mgb} \langle \mbox{Fe} \rangle)^{1/2}$.
So in Fig.~\ref{age3607} we confront [MgFe] to H$\beta$ and compare
our azimuthally averaged data for NGC~3607 with two sets of models --
for [Mg/Fe]$=0$ and for [Mg/Fe]$=+0.3$. As it has been promised by our choice of
the metal-line index, the age estimates are robust to the varying choice of the model [Mg/Fe]; 
they are equal to about 10-12 Gyr. The nucleus is again
outstanding, probably, due to the Balmer emission concentration in
the very center of the galaxy and obvious H$\beta$ emission contamination.
But beyond the nucleus the age gradient along the radius is non-detectable:
at the diagram `H$\beta$, [MgFe]' the index variations occur along the
model sequence of equal age, namely, of $T=10 \pm 2$ Gyr. We may suspect
that the real mean age of the stellar population in 
NGC~3607 is slightly  lower than 10--12~Gyr, because the ionized-gas disk
is known to extend up to $R\approx 15\arcsec$ \citep{shields,macetal} and
therefore the Balmer emission may contaminate the absorption-line H$\beta$
index in some degree all the way. However, our measurements of 
the [OIII]$\lambda$5007 emission-line equivalent width in the nucleus and
in the summed spectrum of  its nearest outskirts ($R=4\arcsec-7\arcsec$)
have given 0.7~\AA\ for the former and 0~\AA\ for the latter, so the effect 
of contamination of  the H$\beta$ index  by emission must be
strong only for the nucleus.  We try to correct it here
by using the measurements of the H$\alpha$-emission equivalent width by
\citet*{hofil3}. To correct the measured Lick 
index H$\beta$ for the emission  we use the well-known fact that
the emission line H$\alpha$ is always much stronger than the H$\beta$ and
the absorption line H$\alpha$ is always weaker than the H$\beta$ one, 
so the equivalent width of the emission line H$\alpha$ can be measured more 
precisely than that of H$\beta$. \citet{hofil3} obtained
$EW(\mbox{H}\alpha \,emis)= 2.03$~\AA\ for NGC~3607 by
subtracting pure-absorption template from the aperture nuclear spectra. 
Let us to note here
that in the nucleus of NGC~3608, another our target galaxy, they saw only
a marginal H$\alpha$ emission, less that $EW=0.4$~\AA,
and no emission lines in the green spectral range.
The well-established and minimum possible intensity ratio
H$\alpha /\mbox{H}\beta$ is known for the radiative excitation of gas by OB-stars
(`HII-region'-type excitation); it is 2.5--2.7.   For shock excitation of the gas 
it is much larger. \citet{hofil3} have classified the nuclear emission
spectrum of NGC~3607 as a LINER,  not a 'HII-type' nucleus. 
\citet{sts2001} have analysed a large heterogeneous
sample of integrated spectra of galaxies of various morphological types 
and have found a good correlation 
$EW(\mbox{H}\beta \, emis)=0.25 EW(\mbox{H}\alpha \, emis)$;
we use  just this relation to calculate $EW(\mbox{H}\beta \, emis)$, which is
in fact the correction of the Lick index H$\beta$ for the emission. 
After it has been corrected for the emission contamination of the H$\beta$ 
index, the nucleus of NGC~3607 has settled to the age sequence of 12~Gyr
in the Fig.~\ref{age3607}, and any age difference between the nucleus and
its outskirts has disappeared. We would like to stress that any 
significant age break is also absent at the border between 
the magnesium-enhanced core of NGC~3607 and 
its `bulge', at $R=5\arcsec - 6\arcsec$. 

\subsection{NGC 3608}

\begin{figure*}
\centering
  \includegraphics[width=17cm]{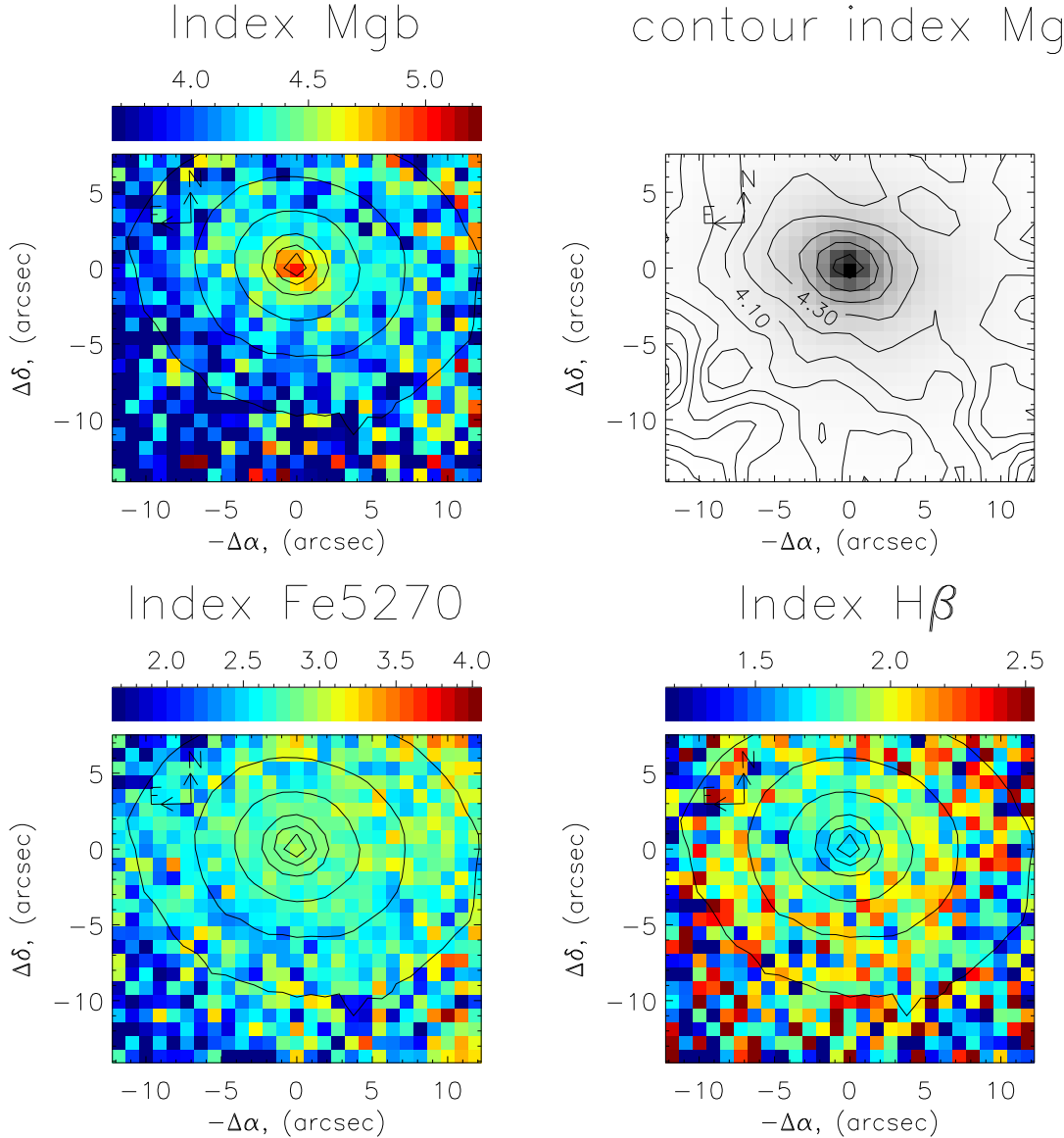}
  \caption{The SAURON Lick index maps for NGC~3608. The green
  ($\lambda$5000~\AA ) continuum is overlaid by isophotes in
  three coloured plots. At the right top the Mgb index distribution
  smoothed strongly with the 2D Gaussian of $FWHM=3\farcs 5$ is plotted
  by isolines to show an orientation of the chemically decoupled
  structure; here the green continuum is gray-scaled}
  \label{imap3608}
\end{figure*}

Figure~\ref{imap3608} presents the Lick index maps for NGC~3608,
similar to those presented for NGC~3607 in Fig.~\ref{imap3607},
but calculated from the SAURON data, so instead of the combined iron index
$\langle \mbox{Fe} \rangle \equiv$(Fe5270+Fe5335)/2
we give here only the Fe5270 surface distribution. We can note a
qualitative resemblance of the index distributions in both galaxies:
just as we have seen in NGC~3607, in its elliptical neighbour the
magnesium index demonstrates a prominent extended peak in the center, and
the iron index shows a flat, homogeneous distribution.
However, the isolines of the smoothed
Mgb distribution (Fig.~\ref{imap3608}, top right) are elongated
in $PA\approx 75\degr$, or along the global major axis of
NGC~3608 (see Table~1), conversely to NGC~3607 where
the magnesium isolines are aligned with the minor axis
of the isophotes. The ellipticity of the
Mgb isocontours within $R\approx 5\arcsec$ is very high, 0.33, 
which value exceeds the isophote ellipticity anywhere
in the galaxy, $1-b/a \leq 0.2$.  Recently the SAURON team has
published their own results of the reduction of the same data \citep{sau6}.
Though not quantified in detail, these results have also implied
the extended magnesium-enhanced region in the center of NGC~3608
with the ellipticity of Mgb-isolines higher than the ellipticity of the isophotes.

\begin{figure}
 \resizebox{\hsize}{!}{\includegraphics{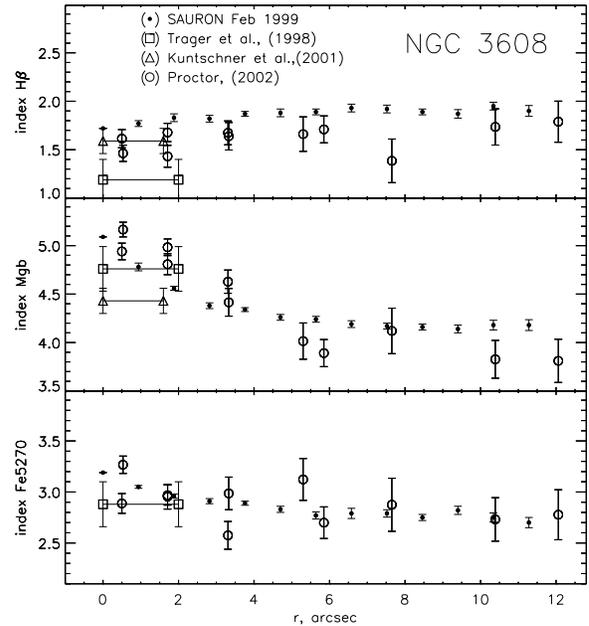}}
 \caption{The SAURON azimuthally averaged Lick index profiles 
for NGC~3608 in comparison with the literature data. The minor-axis long-slit
 data of Proctor (2002) are plotted. The aperture measurements by Trager 
 et al.(1998) 
 and by Kuntschner et al. (2001) are also plotted as connected squares 
 and triangles at $R=0\arcsec -2\arcsec$.}
 \label{indcomp2}
\end{figure}

In Fig.~\ref{indcomp2} we present azimuthally averaged Lick index profiles 
and their comparison to the literature data. 
The shapes of the profiles are qualitatively similar to those in
NGC~3607 (Fig.~\ref{indcomp1}). Perhaps, the magnesium-enhanced core
is somewhat more compact: we would put its border at $R\approx 5\arcsec$.
The break of the Mgb-profile slope at this radius is clearly seen:
between $R=0\arcsec$ and $R=4\farcs 7$ the Mgb index changes by 0.8~\AA,
whereas beyond $R=5\arcsec$ approximation of the azimuthally-averaged
measurements by a linear fit gives the slope of $-0.011\pm 0.010$~\AA\ per
arcsec, or almost negligible, and an intercept of $4.27\pm 0.008$~\AA\ --
compare to the measured central Mgb value of 5.09~\AA. Let us to remind
that the maximum rotation velocity of the counterrotating core
is also achieved near $R\approx 4\arcsec$
\citep{js88}, so in the case of NGC~3608 we deal with
a dynamically \emph{and} chemically decoupled core, just as in the case
of another elliptical galaxy NGC~4365 \citep{bs92}. The Fe5270 measurements
are in good agreement with the minor-axis long-slit data of \citet{prdiss}
and with central aperture data of \citet{trager}, so we may conclude
that in the SAURON data of February 1999 for NGC~3608
the systematic offset of the Fe5270 calibration by 0.4~\AA\ seen in
the later data, e.g. for NGC~3384 \citep{sau2,we2003},
is probably absent. However, there may be a small systematic
shift, by some $+0.2$~\AA, of the H$\beta$ calibration. 

\begin{figure}
 \resizebox{\hsize}{!}{\includegraphics{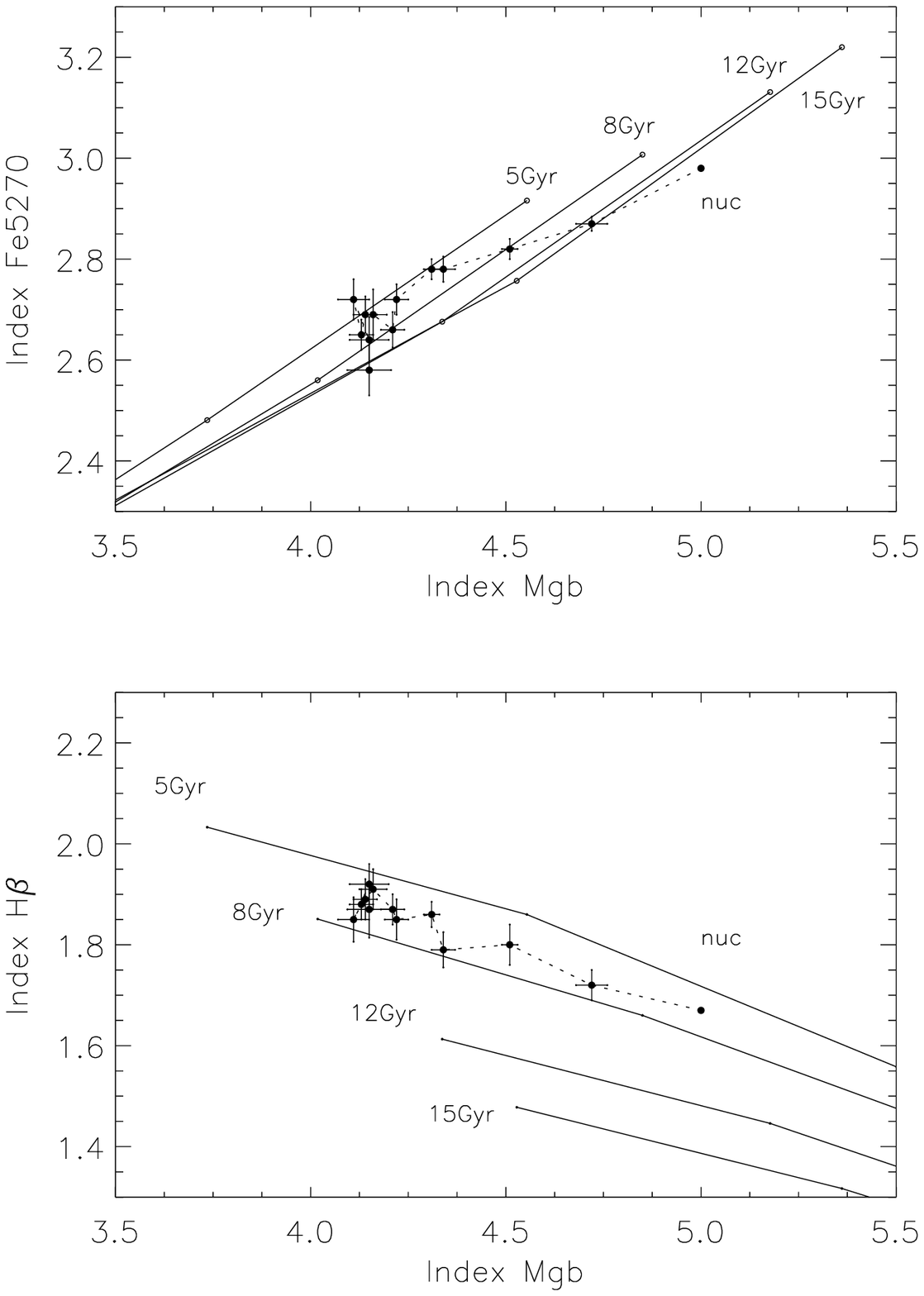}}
 \caption{`Index-index' diagnostic diagrams for the azimuthally averaged
  Lick indices in the center of NGC~3608 taken along the radius with
 the step of $0\farcs 94$ (solid points connected by a dashed line,
 the nucleus being marked by `nuc');
 \emph{(top)} {\bf Fe5270 vs Mgb} diagram, with the models of Thomas et al.
 (2003) for [Mg/Fe]$=+0.3$, and \emph{(bottom)} {\bf H$\beta$ vs
 Mgb}, with the same model set. Small signs in the top plot connected 
 by solid lines belong to stellar population sequences of equal ages,
 particularly, of 5 Gyr, 8 Gyr, 12 Gyr, and 15 Gyr from top to bottom;
 their metallicities are +0.35 and 0.00,
 if one takes the signs from the right to the left}
 \label{ii3608}
\end{figure}

With these azimuthally averaged Lick index profiles, we would try to estimate
radial variations of the mean characteristics of stellar population in the
center of NGC~3608 by using index--index diagrams.
In Fig.~\ref{ii3608} ({\it top})
we compare the SAURON  index radial variations with the models
of \citet{thomod} for [Mg/Fe]$=+0.3$ at the diagram `Fe5270, Mgb'.
The agreement looks rather good:  both chemically
decoupled core and its outskirts have [Mg/Fe]$\approx +0.2 - +0.3$. So to
determine a mean stellar age in this galaxy, we must take the models
with just this magnesium overabundance. Figure~\ref{ii3608} ({\it bottom})
shows the comparison of the azimuthally averaged Lick index data for
NGC~3608 with the models of \citet{thomod} for [Mg/Fe]$=+0.3$ at the
diagram `H$\beta$, Mgb'. Again, as in the case of NGC~3607, we do not
detect any age gradient along the radius: the data sequence follows the
model sequence of equal age, this time it lies between the model age
sequences of $T=5$ Gyr and $T=8$ Gyr. Since the emission lines are very weak
in the center of NGC~3608 (in particular, we have failed to detect noticeable
[OIII]$\lambda$5007 emission line in the green SAURON spectra),
we think that this age estimate, $T=5 - 7$ Gyr, is close to the real
value. Similarly to NGC~3607, NGC~3608 does not reveal
any age break at the transition point from the
magnesium-enhanced core to the rest of the galaxy. The moderate metallicity
difference of $\Delta \log Z \approx +0.2$ is however clearly detected
between the nucleus, $R=0\arcsec$, and the `non-core' stellar body,
$R > 4\arcsec$ (Fig.~\ref{ii3608}, {\it bottom}).

Let us to note that if the systematic shift of the H$\beta$ calibration
by some $0.2$~\AA\ is real, this would result in the age underestimation
by some 4~Gyr.  For example,  \citet{prdiss} has found for the center of NGC~3608 
the mean stellar age of $\sim 9$ Gyr, although the abundance characteristics,
[Mg/Fe]$\approx +0.2 - +0.3$ and [m/H]$=+0.4$ in the nucleus and $+0.2$
in the `bulge', given by him coincide with our estimates.
Similarly, if for NGC~3607 the H$\beta$ index calibration
by \citet{fish96} and \citet{prdiss} is more correct than ours, it would
mean that we overestimate the mean stellar age in the center of NGC~3607
by the same 4~Gyr. So we do not insist on the absolute values of the age
estimates obtained here. The main result of this consideration is that
the relative ages do not show noticeable variations
along the radii within some $10\arcsec$ from the nuclei in both galaxies, 
and particularly, that there are no age drops or rises at the borders of 
the magnesium-enhanced cores.

\section{Discussion}

After a thorough study of the central regions of the brightest galaxies 
of the Leo~II group, the nearest to the center of the group, NGC~3607
and NGC~3608, we note a kinematically distinct area in the center of 
the former and confirm a presence of the counterrotating core in the latter.  
The presence of the counterrotating stellar core in NGC~3608 had been
reported by \citet{js88}; a fall of the rotation velocity after the circumnuclear 
maximum at $R=6\arcsec$  in NGC~3607 was noted by \citet{fisher97}. Both
kinematically decoupled cores have appeared to be also distinguished
by a higher magnesium absorption-line strength, but do not differ from
their outskirts as concerning Fe5270 ($\langle \mbox{Fe} \rangle$) or
H$\beta$.

The most popular hypothesis of the kinematically decoupled core
origin is a hypothesis of smaller elliptical satellite sinking 
\citep{kor5813,js88,balquin}: the
smaller ellipticals have denser cores that can survive during accretion
onto a giant galaxy. Later, when \citet{bs92} have found that the 
kinematically decoupled core of NGC~4365 is also magnesium-enhanced,
the locations of  the decoupled rotation and the magnesium enhancement being
coincident, they have suggested that the decoupled core of NGC~4365
has been formed in a secondary star formation burst after a dissipative
minor merger and subsequent gas accumulation in a circumnuclear
disk. Taking in mind that NGC~3607 and NGC~3608
resemble NGC~4365 as concerning kinematically decoupled,
magnesium-enhanced cores, for our two galaxies we would however 
reject both hypotheses  on the decoupled core origin mentioned above.
The decoupled cores of NGC~3607 and NGC~3608 cannot
be accreted small ellipticals because
the galaxies have peaks of the stellar velocity dispersion and of the
magnesium-line strength just in the centers of the decoupled cores.
Meatime in the frame of the dissipationless accretion scenario the
satellite light must dominate in the center of the merger product and
so must demonstrate a lower metallicity in the center than that of
the giant hosts, and the velocity dispersion must have a dip \citep{balquin}. 
But the scenario of \citet{bs92}
is also inapplicable: in the frame of the dissipative merger hypothesis, the
formation of metal-enriched stars must proceed in a circumnuclear disk,
and in the cases of NGC~3607 and NGC~3608 we have proved that
the magnesium-enhanced structures are not disks. In both galaxies the
areas of the maximum stellar velocity dispersion are elongated. 
Dynamical simulations show that in a triaxial potential the high velocity 
dispersion areas must have an oval shape and to be aligned with the 
model bars \citep{vd97}. The association of the magnesium-enhanced 
areas in NGC~3607 and NGC~3608 with some ellipsoidal bodies is confirmed 
also by  absence of disk-like rotation, when tracing the photometric major 
axes of these structures. In NGC~3607 the magnesium-enhanced 
circumnuclear feature is elongated directly to the kinematical major 
axes of the stars and of the ionized gas --
evidently, it is orthogonal to the circumnuclear disk major axis.
In NGC~3608 the stellar rotation, though proceeding with the
expected spin orientation, is too slow for the visible ellipticity of the
magnesium-index isolines. If we attribute the Mgb-isoline ellipticity, 0.33,
to a hypothetical circumnuclear disk, then the minimum $v_{max}/\sigma$
ratio appropriate to oblate spheroids or disks would be 0.7 
\citep{ill}, and we have $\sim 0.1$ in the center of the galaxy.
The ionized gas rotates in the polar plane with respect to the stellar
rotation in the very center of the galaxy. So we conclude that the
chemically decoupled cores in NGC~3607 and NGC~3608
look like small magnesium-enriched  triaxial structures with some kinds
of polar disks around them -- only gaseous one in NGC~3608
and stellar-gaseous one in NGC~3607. We think that such
a configuration may be possible if occurs inside a global triaxial spheroid.

Though NGC~3607 is classified as S0, some investigators, e.g. \citet{js88}, 
thought it to be an elliptical capturing its diffuse outer envelope from 
NGC~3608 during their interaction.
Perhaps, there exists a principal difference between formation 
of the decoupled cores in elliptical and disk galaxies. In disk galaxies
the chemically decoupled structures are indeed circumnuclear stellar
disks in the most cases. When we have found such structures in
the lenticular galaxies NGC~1023 \citep{sil99a}
and NGC~3384 \citep{we2003}, we have noted
that they are distinguished by a higher
strength of both magnesium and iron absorption lines, the iron-index
enhancement being spatially more extended than that of the Mgb. We
then concluded that the secondary star formation bursts having produced
the decoupled cores were brief and effective in the unresolved nuclei,
resulting in their increased [Mg/Fe], and prolonged during at least
a few Gyr in the circumnuclear disks. In the centers of NGC~3607
and NGC~3608 which may accrete gas from a 3D quasi-spherical hot halo,
and not from an outer fast-rotating disk nor from a supergiant flat gaseous
ring as it was the case of Leo~I group, there were no perhaps initial
favourable conditions for disk formation. Under spheroidal burst geometry,
its duration was short, and only magnesium enhancement had a time
to be imprinted into the secondary stellar populations.

An interesting detail is the absence of the mean age difference between
the decoupled cores and the main bodies of NGC~3607 and
NGC~3608. In disk galaxies, including the above mentioned
lenticulars NGC~1023 and NGC~3384, the
chemically decoupled cores are always `younger' than the bulges, the mean
(luminosity-averaged) stellar age of the nuclei being 3--7 Gyr versus
10--15 Gyr of the bulges. It implies that the secondary star formation
bursts producing the decoupled cores in the disk galaxies are rather
recent. In NGC~3607 and NGC~3608 the decoupled cores do not seem to be
distinguished by a younger mean age of the stellar population.
\citet{prdiss} who studied radial profiles of the stellar population
properties along the minor axes of NGC~3607 and NGC~3608 with the long-slit
spectra has not found any age breaks at $R=5\arcsec - 6\arcsec$ too.
Finally, \citet{n4365sau} who study
2D distributions of the stellar population properties in the center
of NGC~4365, the elliptical galaxy with kinematically decoupled core,
have not also found any age difference between the core and its 
broad outskirts, both being rather old, $T \geq 12$ Gyr. Such  age
`homogeneity' is in some contradiction with the hypothesis of
the secondary nuclear star formation burst having produced the decoupled
cores. However it can be explained by an older age of the decoupled
cores in elliptical galaxies: as the photometric and spectral evolution
becomes very slow at larger ages, one needs a very high precision of the
measured Lick indices to detect an age difference of a few Gyr when
both decoupled cores and their outskirts are older than, say, 8 Gyr.
When a larger statistics on the ages of decoupled cores in elliptical
galaxies become available, this idea would be checked more carefully.

\section{Acknowledgements}
We thank Dr. A. V. Moiseev of the Special Astrophysical Observatory
of the Russian Academy of Sciencies for supporting
the observations at the 6m telescope.
The 6m telescope is operated under the financial support of
Science Ministry of Russia (registration number 01-43).
During the data analysis we have
used the Lyon-Meudon Extragalactic Database (LEDA) supplied by the
LEDA team at the CRAL-Observatoire de Lyon (France) and the
NASA/IPAC Extragalactic Database (NED) which is operated by the
Jet Propulsion Laboratory, California Institute of Technology,
under contract with the National Aeronautics and Space Administration.
The research is partly based on the data taken from the ING Archive
of the UK Astronomy Data Centre and
on observations made with the NASA/ESA Hubble Space Telescope, obtained
from the data archive at the Space Telescope Science Institute, which is
operated by the Association of Universities for Research in Astronomy,
Inc., under NASA contract NAS 5-26555.
The study of the evolution of galaxies in groups
is supported by the grant of the Russian Foundation for Basic Researches
04-02-16087.

\end{document}